\PassOptionsToPackage{table}{xcolor}
\documentclass[sn-basic,pdflatex]{sn-jnl}

\usepackage{graphicx}
\usepackage{multirow}
\usepackage{amsmath,amssymb,amsfonts}
\usepackage{amsthm}
\usepackage{mathrsfs}
\usepackage[title]{appendix}
\usepackage{xcolor}
\usepackage{textcomp}
\usepackage{manyfoot}
\usepackage{booktabs}
\usepackage{tabularx}
\newcolumntype{R}{>{\raggedleft\arraybackslash}X}
\newcolumntype{C}{>{\centering\arraybackslash}X}
\newcolumntype{L}{>{\raggedright\arraybackslash}X}
\definecolor{rowstripe}{rgb}{0.8627, 0.9098, 0.9804}
\usepackage{listings}
\usepackage{array}
\usepackage{xspace}
\usepackage{url}
\usepackage[capitalise,noabbrev]{cleveref}
\usepackage{tikz}
\usetikzlibrary{arrows.meta,positioning,calc,fit}
\definecolor{ovblue}{HTML}{2A78D6}
\definecolor{ovorange}{HTML}{EB6834}
\definecolor{ovink}{HTML}{0B0B0B}
\definecolor{ovgrid}{HTML}{BFBEB9}

\usepackage[most]{tcolorbox}
\newtcolorbox{finding}[1][Finding]{%
  breakable, enhanced, arc=1.5pt, colback=ovblue!5, colframe=ovblue!65,
  boxrule=0.5pt, left=5pt, right=5pt, top=2.5pt, bottom=2.5pt,
  colbacktitle=rowstripe, coltitle=black,
  fonttitle=\bfseries\footnotesize, title={#1},
  before skip=5pt, after skip=5pt, width=\linewidth}

\newcommand{\EFDD}{{\scshape{}EFDD}\xspace}
\newcommand{\FORECAST}{{\scshape{}Forecast}\xspace}
\newcommand{\SFLKIT}{{\scshape{}SFLKit}\xspace}
\newcommand{\TESTSFOURPY}{{\scshape{}Tests4Py}\xspace}
\newcommand{\BUGSINPY}{{\scshape{}BugsInPy}\xspace}
\newcommand{\AMPLE}{{\scshape{}Ample}\xspace}
\newcommand{\FLUCCS}{{\scshape{}FLUCCS}\xspace}
\newcommand{\COMBINEFL}{{\scshape{}CombineFL}\xspace}
\newcommand{\DEEPFL}{{\scshape{}DeepFL}\xspace}
\newcommand{\SCIKITLEARN}{{\scshape{}scikit-learn}\xspace}
\newcommand{\TARANTULA}{\textsc{Tarantula}\xspace}
\newcommand{\OCHIAI}{\textsc{Ochiai}\xspace}
\newcommand{\DSTAR}{\textsc{DStar}\xspace}
\newcommand{\NAISHTWO}{\textsc{Naish2}\xspace}
\newcommand{\GPTHIRTEEN}{\textsc{GP13}\xspace}
\newcommand{\EXAM}{EXAM\xspace}
\newcommand{\DAIKONNAME}{{\scshape{}Daikon}\xspace}
\newcommand{\RQ}[1]{RQ#1}

\expandafter\def\csname fflnum@adaptivetopone\endcsname{0.149}
\expandafter\def\csname fflnum@aggmaxtop1\endcsname{0.161}
\expandafter\def\csname fflnum@aggmeantop1\endcsname{0.165}
\expandafter\def\csname fflnum@aggsumtop1\endcsname{0.162}
\expandafter\def\csname fflnum@alttopone\endcsname{0.163}
\expandafter\def\csname fflnum@augpatchexamavg\endcsname{0.228}
\expandafter\def\csname fflnum@augpatchtop1\endcsname{0.161}
\expandafter\def\csname fflnum@basepatchexamavg\endcsname{0.237}
\expandafter\def\csname fflnum@basepatchtop1\endcsname{0.153}
\expandafter\def\csname fflnum@basepatchtop10\endcsname{0.473}
\expandafter\def\csname fflnum@bestformula\endcsname{Ochiai}
\expandafter\def\csname fflnum@bestformulad\endcsname{0.07}
\expandafter\def\csname fflnum@detectable\endcsname{0.16}
\expandafter\def\csname fflnum@dfunctionexam\endcsname{-0.04}
\expandafter\def\csname fflnum@dhunktop1\endcsname{0.07}
\expandafter\def\csname fflnum@dhunktop10\endcsname{0.09}
\expandafter\def\csname fflnum@dimpacttop1\endcsname{-0.16}
\expandafter\def\csname fflnum@dpatchwastedavg\endcsname{-0.02}
\expandafter\def\csname fflnum@dzpatchwastedavg\endcsname{-0.10}
\expandafter\def\csname fflnum@failfirsttopone\endcsname{0.152}
\expandafter\def\csname fflnum@faminteraction\endcsname{<0.001}
\expandafter\def\csname fflnum@fractionalshare\endcsname{19.4}
\expandafter\def\csname fflnum@headroompoints\endcsname{34.1}
\expandafter\def\csname fflnum@headroomshare\endcsname{2}
\expandafter\def\csname fflnum@hybridtopone\endcsname{0.151}
\expandafter\def\csname fflnum@lambdaattmax\endcsname{0.157}
\expandafter\def\csname fflnum@lambdaattmin\endcsname{0.167}
\expandafter\def\csname fflnum@lambdamax\endcsname{0.55}
\expandafter\def\csname fflnum@lambdamin\endcsname{0.15}
\expandafter\def\csname fflnum@ltrfulldegradeexam\endcsname{80}
\expandafter\def\csname fflnum@ltrfullensnsuperior\endcsname{4}
\expandafter\def\csname fflnum@ltrfullexam\endcsname{0.080}
\expandafter\def\csname fflnum@ltrfullimproveexam\endcsname{206}
\expandafter\def\csname fflnum@ltrfullnsuperior\endcsname{1}
\expandafter\def\csname fflnum@ltrfullseedmajority\endcsname{4}
\expandafter\def\csname fflnum@ltrfullseeds\endcsname{6}
\expandafter\def\csname fflnum@ltrfulltopone\endcsname{0.290}
\expandafter\def\csname fflnum@ltrfulltopten\endcsname{0.679}
\expandafter\def\csname fflnum@ltrgridconfigs\endcsname{14}
\expandafter\def\csname fflnum@ltrgridordered\endcsname{14}
\expandafter\def\csname fflnum@ltrtracecoveragemin\endcsname{94}
\expandafter\def\csname fflnum@ltrvectopone\endcsname{0.274}
\expandafter\def\csname fflnum@medianconstant\endcsname{54.3}
\expandafter\def\csname fflnum@medianfailing\endcsname{1}
\expandafter\def\csname fflnum@medianfeatures\endcsname{62,425}
\expandafter\def\csname fflnum@medianimportancereach\endcsname{4.6}
\expandafter\def\csname fflnum@mediannonzero\endcsname{191}
\expandafter\def\csname fflnum@medianruntime\endcsname{160.4}
\expandafter\def\csname fflnum@naishd\endcsname{-0.29}
\expandafter\def\csname fflnum@nconverged\endcsname{20}
\expandafter\def\csname fflnum@negdensityrho\endcsname{0.97}
\expandafter\def\csname fflnum@negformula\endcsname{Naish2}
\expandafter\def\csname fflnum@negshare\endcsname{26.3}
\expandafter\def\csname fflnum@negsubjects\endcsname{305}
\expandafter\def\csname fflnum@negtop1\endcsname{0.173}
\expandafter\def\csname fflnum@nmodels\endcsname{26}
\expandafter\def\csname fflnum@nmoderatorexpected\endcsname{2.0}
\expandafter\def\csname fflnum@nmoderatorholm\endcsname{0}
\expandafter\def\csname fflnum@nmoderatorsig\endcsname{4}
\expandafter\def\csname fflnum@nmoderatortests\endcsname{40}
\expandafter\def\csname fflnum@nounreached\endcsname{3}
\expandafter\def\csname fflnum@npairedsuperior\endcsname{0}
\expandafter\def\csname fflnum@npositiveprojects\endcsname{9}
\expandafter\def\csname fflnum@nprojects\endcsname{20}
\expandafter\def\csname fflnum@nreachable\endcsname{307}
\expandafter\def\csname fflnum@nsubjects\endcsname{310}
\expandafter\def\csname fflnum@othersdmax\endcsname{0.13}
\expandafter\def\csname fflnum@othersdmin\endcsname{0.09}
\expandafter\def\csname fflnum@permtop1\endcsname{0.143}
\expandafter\def\csname fflnum@posnpairedsuperior\endcsname{5}
\expandafter\def\csname fflnum@postop1\endcsname{0.494}
\expandafter\def\csname fflnum@psinferior\endcsname{9.7}
\expandafter\def\csname fflnum@pssuperior\endcsname{6.9}
\expandafter\def\csname fflnum@rfdegradeexam\endcsname{151}
\expandafter\def\csname fflnum@rfdensityrho\endcsname{0.38}
\expandafter\def\csname fflnum@rfimproveexam\endcsname{74}
\expandafter\def\csname fflnum@rfmaxabsd\endcsname{0.08}
\expandafter\def\csname fflnum@rfmaxabsdz\endcsname{0.17}
\expandafter\def\csname fflnum@rhonegative\endcsname{-0.993}
\expandafter\def\csname fflnum@rhopositive\endcsname{1.000}
\expandafter\def\csname fflnum@seededjaccard\endcsname{0.04}
\expandafter\def\csname fflnum@shareonefailing\endcsname{72}
\expandafter\def\csname fflnum@sharevalue\endcsname{89.8}
\expandafter\def\csname fflnum@signtestp\endcsname{1.000}
\expandafter\def\csname fflnum@topinteraction\endcsname{=0.912}
\expandafter\def\csname fflnum@totalruntime\endcsname{221.9}
\expandafter\def\csname fflnum@wilexamnreject\endcsname{4}
\expandafter\def\csname fflnum@worstformula\endcsname{GP13}
\expandafter\def\csname fflnum@worstformulad\endcsname{0.00}
\expandafter\def\csname fflnum@zerosusp\endcsname{20.8}
\expandafter\def\csname fflnum@zerosuspmax\endcsname{35.2}
\expandafter\def\csname fflnum@zerosuspworst\endcsname{DStar}

\newcommand{\num}[1]{\ifcsname fflnum@#1\endcsname\csname fflnum@#1\endcsname\else\PackageError{numbers}{Unknown number #1}{Regenerate numbers.tex}\fi}
\usepackage{wrapfig}

\begin{document}

\title[How do Execution Features Improve Statistical Fault Localization?]{How do
Execution Features Improve Statistical Fault Localization?}
\subtitle{An Empirical Study}

\author*[1]{\fnm{Marius} \sur{Smytzek}}\email{marius.smytzek@cispa.de}
\author[1]{\fnm{Andreas} \sur{Zeller}}\email{zeller@cispa.de}

\affil[1]{\orgname{CISPA Helmholtz Center for Information Security},
\city{Saarbr\"ucken}, \country{Germany}}

\abstract{
    \textbf{Context.} Statistical fault localization (SFL) ranks suspicious lines from pass and fail coverage spectra, which leave out the data flow, values, and conditions behind a failure.
    \textbf{Objective.} In a registered study, we test whether weighting SFL with the importances of execution features improves localization.
    \textbf{Method.} For all \num{nsubjects} reproducible \TESTSFOURPY{} bugs, we train a random forest per subject on execution features, map its importances to source lines, and weight five SFL formulas with them.
    We evaluate top-$k$, \EXAM{}, and wasted effort with paired effect sizes, a mixed-effects model, and positive and negative controls.
    \textbf{Results.} The weighting barely moves the ranking (top-1 accuracy from \num{basepatchtop1} to \num{augpatchtop1}).
    No paired effect size reaches $0.2$ ($|d_z|\le\num{rfmaxabsdz}$), the method is superior for none of the five formulas, and neither registered hypothesis is supported.
    An oracle weighting is superior for every formula, while random importances do as well as learned ones.
    Turning the tables, a ranker trained on the faults of other projects raises top-1 to \num{ltrfulltopone} and lowers \EXAM{} from \num{basepatchexamavg} to \num{ltrfullexam}.
    Under the paired effect size, it is superior for every formula, while the same ranker with random importances is not.
    \textbf{Conclusions.} Execution features capture how a failing run differs. With a single failing run, as in \num{shareonefailing}\% of subjects, multiple features separate it equally well, and not all of them point to where the fix applies.
    Execution features do improve localization once the knowledge of how to weigh them is learned across faults.
}


\maketitle

\section{Introduction}\label{sec:intro}

Debugging consumes a significant portion of development effort, requiring developers to identify root causes across thousands of lines of code.
Automated fault localization addresses this by ranking the code locations most likely responsible for an observed failure.
Statistical fault localization (SFL) techniques such as \TARANTULA{} \citep{jones2002tarantula} correlate the execution of code lines with test outcomes and are widely adopted because code locations are directly actionable for developers.

SFL is nevertheless limited by the poverty of its input.
A coverage spectrum records only whether a line was executed.
It cannot distinguish among the many lines that every failing test executes.
Passing tests that also execute such a line lower its score, even when the line is faulty.
The ranked list also explains nothing about the conditions under which the failure arises.
Our prior work on Execution-Feature-Driven Debugging (\EFDD{}) showed that execution characteristics beyond line coverage correlate more strongly with failures than line execution alone \citep{smytzek2026efdd}.
Examples are definition-use pairs \citep{santelices2009defuse}, variable-value predicates, and scalar-pair relations \citep{liblit2005statisticaldebugging}.

This raises a question that \EFDD{} itself does not answer.
Does the stronger diagnostic signal of execution features \emph{transfer} to line-level ranking?
Diagnosis and localization are different tasks.
A feature can separate passing from failing runs perfectly and still point at a symptom rather than a cause.
Additionally, the location where the program state turns from sane to infected may be different from the location that a fix needs to target.
In contrast, a feature that has weak discriminability may be located exactly at the fault.
The transfer of features to code lines also passes through a heuristic step, i.e., mapping it back to the source lines it originates from, whose cost is unknown.

\begin{figure*}[t]
\centering
\resizebox{\textwidth}{!}{%
\begin{tikzpicture}[
  font=\footnotesize,
  >={Stealth[length=5pt,width=3.5pt]},
  stage/.style  = {draw=ovgrid, line width=0.5pt, rounded corners=2.5pt,
                   align=center, inner sep=4pt, minimum height=10mm,
                   text width=24mm, fill=white,
                   execute at begin node={\hyphenpenalty=10000\exhyphenpenalty=10000}},
  addstage/.style = {stage, draw=ovblue!70},
  blend/.style  = {circle, draw=ovorange, line width=0.9pt, fill=ovorange!10,
                   inner sep=0pt, minimum size=8mm},
  gate/.style   = {circle, draw=ovorange, line width=0.9pt, fill=white,
                   inner sep=0pt, minimum size=7mm},
  result/.style = {stage, draw=ovink!65, fill=ovink!4, text width=22mm},
  flow/.style   = {->, draw=ovink!50, line width=0.8pt},
  addflow/.style= {->, draw=ovblue!80, line width=0.9pt},
  lane/.style   = {font=\footnotesize\itshape, inner sep=0pt},
]
\def\dx{30mm}
\def\bandl{0.56*\dx}
\def\bandr{3.70*\dx}

\fill[ovblue!7, rounded corners=4pt] (\bandl, 1.15) rectangle (\bandr, 2.35);
\fill[ovink!5,  rounded corners=4pt] (\bandl,-0.85) rectangle (2.62*\dx, 0.35);

\node[lane, text=ovblue, anchor=south west] at (\bandl, 2.44)
  {added by this study};
\node[lane, text=ovink!60, anchor=north west] at (\bandl,-0.94)
  {baseline SFL, unchanged};

\node[stage, text width=20mm] (runs) at (0,0.75)
  {passing and failing test executions};

\node[addstage] (vec)  at (1.14*\dx,1.75) {one feature vector per execution};
\node[addstage] (rf)   at (2.13*\dx,1.75) {random forest importances $I(f)$};
\node[addstage] (wmap) at (3.12*\dx,1.75) {per-line weight $w(l)\in[\varepsilon,1]$};

\node[stage] (spec) at (1.14*\dx,-0.25) {line spectra $e_f,\;e_p$};
\node[stage] (susp) at (2.13*\dx,-0.25) {suspiciousness $\mathrm{susp}'(l)$};

\node[gate]   (mult) at (4.16*\dx,1.75) {$\otimes$};
\node[blend]  (plus) at (4.85*\dx,0.75) {$\oplus$};
\node[result] (rank) at (5.84*\dx,0.75) {ranked lines, ties broken at random};

\draw[addflow] (runs.north) |- (vec.west);
\draw[flow]    (runs.south) |- (spec.west);
\draw[addflow] (vec)  -- (rf);
\draw[addflow] (rf)   -- (wmap);
\draw[flow]    (spec) -- (susp);
\draw[addflow] (wmap.east) -- (mult.west);
\draw[flow] (susp.north) -- ++(0,0.5) -| (mult.south);
\draw[addflow] (mult.east) -| (plus.north)
  node[pos=0.75, right=1pt, text=ovblue, font=\scriptsize] {$\lambda$};
\draw[flow] (susp.east) -| (plus.south)
  node[pos=0.75, right=1pt, font=\scriptsize] {$1-\lambda$};
\draw[flow] (plus) -- (rank);

\node[font=\scriptsize, text=ovink!75, anchor=north] at (2.6*\dx,-1.50)
  {$\mathrm{susp}(l)=(1-\lambda)\,\mathrm{norm}\bigl(\mathrm{susp}'(l)\bigr)
     +\lambda\,\mathrm{norm}\bigl(\mathrm{susp}'(l)\cdot w(l)\bigr)$,\quad
   $\lambda=0.35$};
\end{tikzpicture}%
}
\caption{The registered approach.
The underlying fault localization remains intact (gray lane).
Execution-feature evidence inputs a weight per line (blue lane).
Hence, our study measures a controlled paired delta instead of a new algorithm.
The weight multiplies the suspiciousness ($\otimes$) before the blend ($\oplus$), so a line that the formula scores at zero stays at zero for every weight.
\Cref{sec:ceiling} measures what that costs.}\label{fig:overview}
\end{figure*}
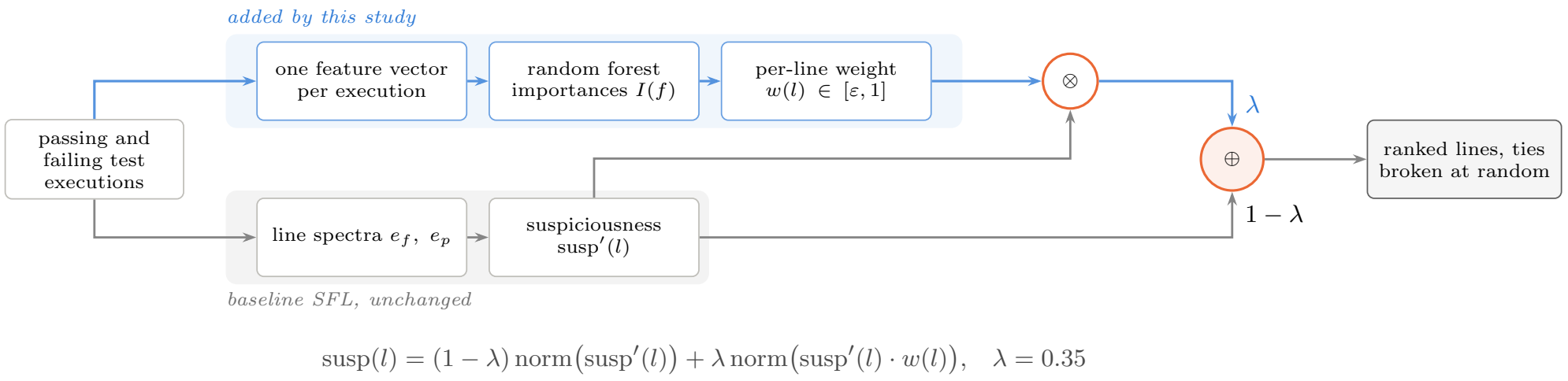

We ran a pre-registered study \citep{nosek2018preregistration} with a deliberately conservative design.
We keep each established SFL formula intact and only \emph{modulate} it with a per-line weight derived from execution-feature importances (\Cref{fig:overview}).
Because the weight modulates the established score instead of replacing it, the design isolates the contribution of execution features from every other design choice.
As \Cref{sec:ceiling} shows, this conservatism carries a cost that we did not foresee at registration.
The protocol, including its techniques and metrics, was registered at the ICSME~2026 Registered Reports track \citep{smytzek2026ffl} before any outcome was computed.
This manuscript documents its complete evaluation and states each deviation from the protocol where it applies.

The result we found: the transfer does not happen.
Weighting the five formulas by the random-forest importances, as registered, moves top-1 accuracy from \num{basepatchtop1} to \num{augpatchtop1}.
Every primary outcome changes in the beneficial direction.
However, no change is large enough to count as an improvement under the registered decision rule.
The method is superior for no formula, and neither registered hypothesis is supported.
An oracle that may express perfect knowledge only through the same mapping and blend reaches top-1 \num{postop1}, so the pipeline itself can carry a signal and is, therefore, not the reason for the results.
The learned importances do not supply it, since uniformly random importances do at least as well (top-1 \num{negtop1}) and permuting the learned importances across features changes little (top-1 \num{permtop1}).
In \num{shareonefailing}\% of subjects, a single failing test is all there is to learn from.
Multiple features may separate it from the passing runs equally well, and which of them a random forest credits is close to arbitrary.
Even though many of these features point at a fault-revealing location, causes, or symptoms, it is nearly unattainable to pinpoint the location fixed by the reference patch.

A single failing run cannot reveal which execution features mark the faulty lines, even though such features are known to help with diagnosis \citep{smytzek2026efdd} and repair \citep{smytzek2026forecast}.
However, the faults of other projects can guide the localization.
We therefore turn the tables and train a ranker on the SFL scores, the feature weightings, and signals from ordered execution traces.
Every project is ranked by a model trained only on the other projects, e.g., the model for a \texttt{matplotlib} bug is only trained on the bugs excluding all \texttt{matplotlib} bugs.
We evaluate it with the registered outcomes and decision rule.
The ranker raises top-1 accuracy from \num{basepatchtop1} to \num{ltrfulltopone} and top-10 from \num{basepatchtop10} to \num{ltrfulltopten}, and it lowers \EXAM{} from \num{basepatchexamavg} to \num{ltrfullexam}.
Under the paired effect size, it is superior for every formula.
The same ranker without the features, or with random importances in their place, is not.
Execution features do improve statistical fault localization, then, but through evidence learned across faults rather than from the one failing run at hand.

\paragraph{Contributions.} Overall, this study makes the following contributions to the empirical understanding of execution features and statistical fault localization:
\begin{description}
  \item[A pre-registered evaluation.] We evaluate execution-feature-weighted SFL on all \num{nsubjects} reproducible \TESTSFOURPY{} bugs and five SFL formulas, following the registered protocol.
 The weighting yields no practically relevant improvement (\Cref{sec:results}).
  \item[Outcome-neutral controls.] An oracle, random importances, and permuted importances pass through the same pipeline as the learned ones.
 They show that the pipeline can carry a signal and that the learned importances do not supply it (\Cref{sec:controlresults}).
  \item[An account of the failure.] We explain why the transfer from diagnosis to localization fails.
 Most subjects offer a single failing run, and multiple features then separate it from the passing runs equally well, and not all of them are suited for the needed localization (\Cref{sec:learner}).
  \item[A ceiling of the combination rule.] We measure how much any weighting that the registered rule can express is able to gain.
 A line that the formula scores at zero stays at zero for every weight, which bounds all such methods (\Cref{sec:ceiling}).
  \item[A cross-project ranker.] We train a ranker on the SFL scores, the feature weightings, and signals from ordered execution traces.
 Each project is ranked by a model trained only on the other projects, and under the paired effect size, the ranker is superior for every formula (\Cref{sec:ltr}).
  \item[A replication package.] We release the feature vectors, importances, mappings, traces, and analysis scripts.
 Alternative weightings and rankers can thus be evaluated without re-running the instrumentation.
\end{description}

\section{Background and Related Work}\label{sec:background}

\subsection{Spectrum-based fault localization}

Spectrum-based fault localization ranks program elements by correlating their execution with test outcomes.
\TARANTULA{} made the idea concrete for developers by coloring source lines with a suspiciousness score derived from how often each line appears in failing versus passing runs \citep{jones2002tarantula, jones2005tarantula}.
\AMPLE{} applied the same logic of suspicious elements to method sequences \citep{dallmeier2005ample}.

Most subsequent work has varied the coefficient that turns a spectrum into a score.
\OCHIAI{}, borrowed from molecular biology, proved more accurate than the original \TARANTULA{} ratio on several benchmarks \citep{abreu2006ochiai, abreu2007accuracy, abreu2009practical}.
\DSTAR{} generalizes the numerator to a tunable power \citep{wong2012dstar}, and crosstab formulations apply a statistical independence test to the same four counts \citep{wong2008crosstab}.
Theoretical work has asked which coefficients can be optimal at all.
\citet{naish2011model} derive a model of spectra-based diagnosis and identify formulas that are optimal under stated assumptions, \citet{lee2009naish} study the role of non-redundant test cases, and \citet{landsberg2015measures} evaluate measures and derive an optimizing scheme.
Genetic programming has searched the space of formulas directly and produced coefficients such as \GPTHIRTEEN{} that are competitive with hand-designed ones \citep{xie2013gp, wang2011search}.
These results do not agree on a single winner, and relative performance depends on the subjects and the evaluation protocol \citep{pearson2017evaluating, heiden2019purespectrum}.
We therefore evaluate against five established formulas.
\SFLKIT{} provides the configurable Python implementation of spectra, predicates, and coefficients that this study builds on \citep{smytzek2022sflkit}.

\subsection{Signals and learned rankings}

A coverage spectrum tells only whether an element was executed.
A substantial line of work argues that richer dynamic signals localize better than spectra.
Definition-use coverage captures the data dependence that plain statement coverage misses \citep{santelices2009defuse, ribeiro2019dataflow}.
Statistical bug isolation leverages predicates over the program state and identifies those predicates that are more strongly correlated with the failure \citep{liblit2005statisticaldebugging}.
Later work introduces the propagation of infected states \citep{zhang2009capturing}, the context of method calls \citep{vancsics2021calls, yan2023context}, variable values \citep{jiang2023variable}, and general feature abstractions \citep{lei2022featurefl}.
Predicate-based and spectrum-based signals complement each other \citep{jiang2020combining}.

Another line of research derives behavioral specifications and uses their violation as localization signals.
\DAIKONNAME{} infers likely invariants over program points \citep{ernst2007daikon}.
These invariants can be combined with spectra in a learning-to-rank model \citep{le2016savant} or used to sharpen suspiciousness scores directly \citep{gopinath2012specifications}.
Path-based correlations \citep{le2010faultcorrelations} and program slicing \citep{xu2005slicing, soremekun2021slicing} narrow the search space by following the dependences that lead to the failure instead of correlations.
Our own work on Execution-Feature-Driven Debugging gathers seventeen such signals into a single vocabulary.
It shows that they separate failing from passing \emph{runs} better than line coverage does \citep{smytzek2026efdd}.
This study asks whether that benefit transfers from classifying runs to ranking lines, a question that our previous work does not answer.

Rather than designing a coefficient, many approaches learn one.
PageRank has weighted tests by how much they discriminate and fed the reweighted spectrum into a standard formula \citep{zhang2017pagerank, zhang2021prfl}.
Constrained feature selection over traces improves top-$k$ accuracy \citep{le2015featureselection}, and learning-to-rank models combine dynamic, mutation-based, and static features \citep{kim2019prince}.
Deep models expand this concept to various diagnosis dimensions \citep{li2019deepfl}, learned coverage representations \citep{li2021covrepresentation}, code transformations \citep{li2017transforming}, and transferred knowledge \citep{meng2022transfer}.
Model-agnostic attribution \citep{lundberg2017shap} has also been applied to the ranking itself \citep{widyasari2022xai4fl}.

These methods learn the ranking and, in effect, replace the suspiciousness formula with a model.
The registered design instead keeps each formula intact and uses a random forest \citep{breiman2001randomforests} only as a per-subject attribution device.
What it adds is an inspectable weight rather than a new ranker.
The study therefore measures a controlled paired delta.
When we turn the tables in \Cref{sec:ltr}, we return to this family and learn across projects, as \FLUCCS{} \citep{sohn2017fluccs}, \COMBINEFL{} \citep{zou2021combinefl}, and \DEEPFL{} \citep{li2019deepfl} do.

\subsection{Evidence, evaluation and benchmarks}

A third family changes the evidence rather than the inference.
Mutation-based localization runs mutants and treats the mutants whose behavior tracks the failure as evidence about location \citep{papadakis2014mutation, papadakis2015metallaxis, hong2015mutation}.
In contrast, other work tries to improve the test suite itself to make it more suitable for fault localization.
For instance, test-case purification splits assertions to sharpen the spectrum \citep{xuan2014purification}.
Entropy-based generation creates tests that maximize diagnostic information \citep{campos2013entbug}.
Prioritization reorders tests for faster localization \citep{yoo2013priorization}.
\citet{ceccato2015testgen} asked whether automatically generated tests help a developer debug, and \citet{just2018developertests} investigated how developer-provided and user-provided tests differ.
This family matters for our findings, because the median subject offers a single failing execution, and these are the techniques that attack that scarcity.

\citet{steimann2013threats} catalog threats to the validity of accuracy assessments.
\citet{soremekun2023evaluating} show experimentally how much common assumptions move the reported numbers.
Studies on real faults report markedly weaker performance than studies on seeded ones \citep{pearson2017evaluating, widyasari2022realworld}.
\citet{parnin2011automated} found that a ranked list does not by itself make developers faster, a conclusion revisited a decade later \citep{parnin2021automated}.
Practitioners' expectations of localization tools have been surveyed \citep{kochhar2016practitioners}, and eye-tracking has shown how developers consume rankings \citep{xie2016focustracking}.
Tool designs that present hypotheses rather than rankings follow from the same concern \citep{alaboudi2023hypothesizer}.

\BUGSINPY{} collects reproducible Python faults from real projects \citep{widyasari2020bugsinpy, widyasari2020bugsinpysource}.
\TESTSFOURPY{} builds on it with a uniform harness for checkout, build, instrumentation, and test execution \citep{smytzek2024tests4py}, which makes a study of this size feasible.
\Cref{sec:discussion} reports two properties of this benchmark that bound any line-level evaluation on it.

Automated program repair edits the locations that localization suggests \citep{weimer2009genprog, legoues2012genprog, nguyen2013semfix, mechtaev2016angelix, long2015spr, bader2019getafix} and depends strongly on where it is told to look \citep{qi2015kali, long2016search}.
Repair has in turn been used to evaluate localization \citep{qi2013aprevaluation, lou2020aprsfl}.
Approaches that characterize failure-inducing inputs \citep{kampmann2020alhazen, eberlein2023avicenna, bohme2023learn2fix} or use large language models \citep{kang2024autofl} complement the ranking of lines.

\subsection{Positioning of this study}

It is already established that weighting and learning can improve rankings and that execution features can carry signals that lines cannot explain alone.
However, two gaps remain.
First, it is not known whether feature-based weighting provides robust improvements for SFL baseline techniques on real faults.
Second, it is not known whether such conclusions survive an evaluation that does not rest on a single reference patch.
This study addresses both gaps for the \TESTSFOURPY{} benchmark.
A positive result was uncertain at registration, because the mapping is heuristic, importances may be unstable, and failing tests are sparse.
The study measures a \emph{controlled paired delta}, the effect of adding execution-feature evidence to established SFL techniques with other variables held fixed.
LLM-based localization \citep{kang2024autofl} would confound this delta with prompt design and model priors, so we leave it for future work.

\section{Study Design}\label{sec:design}

The study follows the registered protocol.
Where the executed study had to deviate from the protocol, we say so at the point where it applies.

\subsection{Research questions and hypotheses}\label{sec:rqs}

\subsubsection{Hypotheses}

We form two hypotheses to measure the influence of execution-feature-weighting on fault localization.

\begin{description}
\item[$H_{\mathrm{ACC}}$ (Accuracy).] Augmenting SFL with execution-feature weights improves fault localization effectiveness over baseline SFL, because execution features capture failure-relevant behavior that line spectra alone cannot represent.
\item[$H_{\mathrm{ROB}}$ (Robustness).] The improvement is robust across different SFL formulas, because execution-feature signals are orthogonal to the suspiciousness formula used by the baseline.
\end{description}

\subsubsection{Research questions}

To verify these hypotheses, we formulate the following research questions:

\begin{description}
\item[\RQ{1} Patch-line accuracy.] Does execution-feature augmentation improve
top-$k$ localization metrics (top-1, top-5, top-10) for reference-patch lines in \TESTSFOURPY{} compared with each baseline SFL technique?
\item[\RQ{2} Developer effort.] Does execution-feature augmentation improve or
maintain line- and function-level inspection-effort metrics compared with each baseline SFL technique?
\item[\RQ{3} Broader ground truth.] Do augmented rankings discover
failure-relevant locations beyond the reference patch, as measured by dynamically discovered test-impact sets?
\item[\RQ{4} Robustness.] Is the improvement direction consistent across all
five baseline techniques (\TARANTULA{}, \OCHIAI{}, \DSTAR{}, \NAISHTWO{}, and \GPTHIRTEEN{}) and, exploratorily, across the implemented ranking variants?
\end{description}

\subsection{Variables}

\subsubsection{Independent variables}

We vary two factors: the localization method and the baseline formula (\TARANTULA{}, \OCHIAI{}, \DSTAR{}, \NAISHTWO{}, and \GPTHIRTEEN{}).
The localization methods are the unmodified SFL baselines, the primary random-forest feature weighting, and a separate exploratory family of ranking variants (\Cref{sec:variants}).

\subsubsection{Dependent variables}

\emph{Patch-line outcomes} (\RQ{1} for top-$k$, \RQ{2} for effort) are top-1, top-5, and top-10 accuracy, the \EXAM{} score, and wasted effort (WE in the tables) in the best, average, and worst case, all against the reference patch lines.
\emph{Function-level outcomes} (\RQ{2}) aggregate the ranking to functions, the granularity at which developers read code.
They are the function-weighted mass@1/3/5, the faulty-function probability mass, the function-weighted inspection effort, and the practical effort until a faulty function is reached.
\emph{Test-impact outcomes} (\RQ{3}) use the reference patch hunks and dynamically discovered test-impact sets as targets, with impact-found@1/5/10, the minimum impact rank, and the number of impact locations ranked.
\emph{Exploratory ranking-variant outcomes} (\RQ{4}) apply the same metrics to the variants of \Cref{sec:variants}.
\emph{Diagnostic metrics} are the interpretive quantities of \Cref{sec:diagnostics}, namely importance concentration, entropy, skewness, feature dropping, mapping conflicts, and weak-signal counts.
We also report runtime, timeouts, exclusions, and trace or storage failures as feasibility outcomes.

\subsubsection{Confounding variables}\label{sec:confounders}

The confounders we consider are bug difficulty, the project size, the count of executable lines, the size of the test suite, the number of failing tests, coverage density, and the chosen SFL formula.
First, we need to define bug difficulty.
However, any outcome-derived definition would be circular.
We therefore consider a practical view and define bug difficulty as the size of the reference fix, $\log(1+|\text{reference lines}|)$, because a fix spanning several locations is harder to pin down than a single-line one. 
All other confounding variables are measured exactly as registered.
The project size is the number of lines of code that \TESTSFOURPY{} reports for the project.
The count of executable lines is the number of lines in the line spectrum, i.e., the lines that the localization ranks.
The size of the test suite counts the tests run for the bug, and the number of failing tests counts those among them that fail.
Coverage density is the share of line-test pairs in which the test executes the line.
The SFL formula enters the model as a factor.
We take $\log(1+x)$ of every count and standardize all covariates before fitting the model.

For each bug, baseline and the weighting methods run on identical code, tests, and spectra.
Every time-invariant property of the bug thus takes the same value for both methods and cannot shift the average paired contrast.
These properties still govern how large the effect is and where it concentrates, so the mixed-effects model of \Cref{sec:stats} includes them as covariates (\Cref{sec:whyanalysis}).

\subsection{Dataset}

We evaluate all 310 reproducible \TESTSFOURPY{} bugs \citep{smytzek2024tests4py} to avoid selection bias.
\TESTSFOURPY{} provides real-world Python faults with tests and reference patches.
Each bug is one unit in our analysis.
\Cref{tab:subjects} characterizes the bugs evaluated in this study.

\begin{table}[t]
\caption{Subjects of the study.
Values are per-project medians over the reproducible bugs, and \emph{features} is the number of distinct execution features observed for a subject.}
\label{tab:subjects}
\centering
\footnotesize
\begingroup\scriptsize\setlength{\tabcolsep}{3.2pt}
\begin{tabularx}{\linewidth}{Lrrrrrrr}
\toprule
Project & Bugs & LOC & Exec.\ lines & Tests & Failing & Features & Runtime (s) \\
\midrule
\rowcolor{rowstripe}ansible & 17 & 701,141 & 571 & 54 & 1 & 16,505 & 85 \\
black & 21 & 4,110 & 1,565 & 72 & 1 & 102,069 & 6,938 \\
\rowcolor{rowstripe}calculator & 1 & 20 & 14 & 12 & 4 & 267 & 4 \\
cookiecutter & 3 & 1,100 & 94 & 11 & 3 & 1,693 & 4 \\
\rowcolor{rowstripe}expression & 1 & 108 & 52 & 10 & 2 & 788 & 1 \\
fastapi & 16 & 3,659 & 760 & 14 & 1 & 34,098 & 49 \\
\rowcolor{rowstripe}httpie & 5 & 2,255 & 649 & 19 & 1 & 45,984 & 17 \\
keras & 41 & 22,237 & 2,897 & 19 & 1 & 116,995 & 328 \\
\rowcolor{rowstripe}luigi & 27 & 11,261 & 1,095 & 43 & 1 & 60,940 & 831 \\
markup & 2 & 14 & 13 & 4 & 1 & 267 & 2 \\
\rowcolor{rowstripe}matplotlib & 29 & 64,261 & 7,909 & 17 & 1 & 691,012 & 12,000 \\
middle & 2 & 18 & 9 & 6 & 1 & 201 & 2 \\
\rowcolor{rowstripe}pysnooper & 2 & 342 & 214 & 18 & 3 & 5,198 & 50 \\
sanic & 4 & 4,816 & 1,061 & 28 & 1 & 50,047 & 47 \\
\rowcolor{rowstripe}scrapy & 31 & 11,770 & 492 & 12 & 1 & 8,222 & 9 \\
spacy & 9 & 73,083 & 1,125 & 7 & 1 & 460,193 & 1,185 \\
\rowcolor{rowstripe}thefuck & 31 & 2,256 & 125 & 11 & 2 & 1,178 & 2 \\
tornado & 16 & 11,655 & 2,267 & 36 & 1 & 115,604 & 235 \\
\rowcolor{rowstripe}tqdm & 9 & 1,485 & 541 & 44 & 1 & 36,262 & 40 \\
youtubedl & 43 & 63,545 & 5,341 & 29 & 1 & 808,200 & 720 \\
\rowcolor{rowstripe}\textbf{All} & 310 & 12,034 & 1,182 & 20 & 1 & 62,425 & 160 \\
\bottomrule
\end{tabularx}
\endgroup
\end{table}

\subsection{Definitions}\label{sec:definitions}

Reference faulty lines are the exact lines that the official benchmark fix modifies in its patch files.
Patch-hunk sets contain every line of the buggy revision inside a hunk that the official fix touches.
Every function that contains at least one reference line or patch-hunk line counts as faulty for the function-level evaluation, i.e., a fix spanning several functions yields several faulty functions.
Test-impact sets are derived from the observed line spectra.
A line belongs to the set if at least one failing test observes it and either no passing test observes it or its failing-observation ratio is at least $0.6$. 
If this yields an empty set, all lines observed by a failing test form the set.
A line that was observed in a majority of failing tests is more plausibly failure-relevant.
Lower thresholds admit incidental lines, and higher ones discard relevant lines that only some failing tests exercise.
The registered sensitivity analysis varies the threshold ($\{0.5, 0.6, 0.7, 0.8\}$).
We separate patch-hunk and test-impact sets, so that an improvement against the reference is not interpreted as better coverage of failure-relevant behavior.

\subsection{Outcomes}\label{sec:outcomes}

The protocol names the outcome families but leaves part of their arithmetic to the implementation, which we fix here so that the numbers are reproducible.

\emph{Patch-line outcomes} are computed with \SFLKIT{} and follow its definitions.
Top-$k$ is the probability, over the randomized tie-breaking replicates, that at least one reference line lies within the first $k$ positions.
A tie group that lies entirely above the cut-off is always inside, and only the group that the cut-off splits is broken at random.
\EXAM{} and wasted effort involve no tie breaking.
A location in a tie group takes the midpoint rank of its group.
\EXAM{} is that rank divided by the number of ranked locations, which are the lines the subject's tests execute.
Wasted effort is the number of locations a developer inspects up to and including the target's tie group, so every tied competitor counts as inspected.
A reference line that no test executes is charged the end of the ranking, so it is neither rewarded nor silently dropped, and its \EXAM{} is~1.
The best, average, and worst case take the minimum, the median, and the maximum over the reference lines of a subject, as in \SFLKIT{}.

\emph{Function-level outcomes} score a function by the maximum suspiciousness of its ranked lines and rank functions with the same randomized tie-breaking.
Mass@$k$ is the expected share of the normalized score mass that faulty functions hold within the top $k$, and the faulty-function mass is their total share.
Function-weighted inspection effort is the rank of the first faulty function divided by the number of ranked functions.
Practical effort is the share of ranked function lines read up to and including the first faulty function.

\emph{Test-impact outcomes} report impact-found@$k$ (impact@$k$ for short), the probability that at least one location of the target set is ranked within the top $k$.
They also report recall@$k$, the expected share of the target set inside the top $k$.
The minimum impact rank is the expected rank of the highest-ranked target, and the number of impact locations ranked counts the targets that any test executes.

\subsection{Decision rule}\label{sec:decision}

The primary method is the random-forest feature weighting, compared against each of the five unmodified SFL formulas.
The primary outcomes are the five reference-patch metrics, top-1, top-5, and top-10 accuracy, the \EXAM{} score, and average-case wasted effort.
The protocol states that each is ``computed after randomized tie-breaking and averaged per subject'', which is accurate for top-$k$ only.
\EXAM{} and wasted effort are defined as a deterministic rank of a tie group (\Cref{sec:outcomes}), and we compute them that way.
For a given baseline formula, the augmented method is \emph{superior} if the method coefficient of the confounder-adjusted model improves at least four of the five metrics with a non-negligible standardized effect of at least $0.2$ and degrades none of them by a non-negligible effect.
$H_{\mathrm{ACC}}$ is supported if the random-forest weighting is superior for a majority of the five SFL formulas.
$H_{\mathrm{ROB}}$ is supported if it is superior for all five formulas.
The superiority of $H_{\mathrm{ROB}}$ is determined by the method$\times$formula interaction and a sign test.
The exploratory ranking variants form a family with its own control.
If the primary comparison is negative or mixed, the positive results for the variants only count as exploratory evidence about alternative aggregation and not as support for the primary claim.

The rule is purposely phrased in effect sizes rather than statistical significance ($p$-values). 
With 310 subjects per formula, a paired comparison detects an effect of about $0.16$ at 80\% power \citep{cohen1988power} but may classify trivial differences as statistically significant.
A rule based on $p$-values could let a benefit with little to no practical value be interpreted as support.
Fixing the threshold for effect size of $0.2$ in advance states what would be worth acting on.
Additionally, it makes a null interpretable as a bounded absence instead of an inconclusive result (\Cref{sec:whyanalysis}).
The protocol defines the threshold as $|d|\ge0.2$ but does not name the standardizer.
While running the study, we found that the choice of the standardizer matters.
Hence, we apply the rule with the paired effect $d_z$.
It is the mean per-subject difference between the baselines and the augmented results, divided by the standard deviation of those differences.
This study asks how much our method changes the ranking of the same bug.
Because each subject serves as its own control, effect sizes from a paired design estimate a different parameter than those between independent groups \citep{morris2002combining}.
For evaluations about the change within a subject, $d_z$ is the appropriate effect size \citep{lakens2013calculating}.
The power analysis above already assumes it, since $0.16$ is the effect that a paired test detects with 310 subjects \citep{lakens2013calculating}.
For two independent groups of that size, the detectable effect would be about $0.23$.

As a check, we apply the same rule with the registered $d$, the conventional reading of the protocol.
It divides the method coefficient by the between-subject standard deviation of the baseline outcome for the same formula, so it expresses the improvement in units of how much subjects differ under the baseline.
That spread plays no role in a paired comparison, where every subject serves as its own control.
For outcomes measured in lines, it is also dominated by a few subjects whose fault lies thousands of lines down the ranking, so a change that is consistent across subjects looks small.
The between-subject standardizer is preferred when effect sizes are compared across designs, as in a meta-analysis \citep{dunlap1996metaanalysis, lakens2013calculating}, which this study does not do.
We noticed the problem while running the study, after the first results were known.
For the registered weighting, its variants, and the controls, both standardizers give the same verdict for every formula (\Cref{sec:rq1,sec:controlresults}), so the choice changes no registered conclusion.

\section{Approach}\label{sec:method}

The registered approach extracts execution features, trains a random forest to identify failure-correlated features, maps the feature importances to source lines, and boosts the baseline SFL scores with the resulting weights.

\subsection{Feature extraction}

A \emph{feature} is an execution-derived signal of \EFDD{}, encoded as a binary or ternary variable.
Following the \EFDD{} taxonomy \citep{smytzek2026efdd}, the features come from the coverage, value, condition, and exception families.
They include line, branch, function, and loop execution, definition-use pairs, variable-value and length properties, scalar-pair relations, condition outcomes, and exception exits.
We use \EFDD{} on top of \SFLKIT{} \citep{smytzek2022sflkit} to instrument Python code and to generate one labeled feature vector per passing or failing test execution.
Each vector records the presence and values of these features over an entire run, together with the run's pass or fail label.
One vector is the natural unit for the granularity of a test execution.
A passing and a failing vector suffice to compute feature differences.
Hence, no run needs to be subdivided.
Unobserved features take a fixed default value, which does not need any further assumption.
Features that are constant within a subject, i.e., they have the same value in all executions, are removed before training.
\SFLKIT{} classifies a test that surpasses the 120-second execution budget as undefined instead of passing or failing.
Such tests do not contribute a labeled vector.

\subsection{Execution features}\label{sec:vocabulary}

The whole transfer from diagnosis to localization depends on how a feature is mapped to a source line.
An \EFDD{} feature is a property of a single execution, encoded as a binary variable (present or absent) or a ternary one (true, false, or not observed).
\Cref{tab:vocabulary} lists all seventeen types, their families, and the lines to which each contributes its importance.
Two properties of the vocabulary matter for the results.

First, only two kinds of features map to more than one line.
A definition-use pair contributes to both the defining and the using line.
The construct-level types (branch, function, loop, and exception exit) contribute to every line of the construct they describe.
Every other type maps to the single line at which it is observed.
This is why the mapping-conflict rate is low, and why the alternative mapping can differ only for definition-use pairs (\Cref{sec:mapping}).

Second, all seventeen types are extracted for every subject and enter the forest together, and every feature is binary or ternary.
This avoids the cardinality bias of impurity-based importance on mixed-type data \citep{strobl2007importancebias}, and the data-quality check of \Cref{sec:controls} verifies the encoding on every subject.

\begin{table}[t]
\caption{The execution-feature vocabulary and the lines to which each feature type contributes its importance.
\emph{Construct lines} are all lines of the branch, function, or loop that the feature describes, resolved from the source.}%
\label{tab:vocabulary}
\centering
\footnotesize
\begin{tabularx}{\linewidth}{llLl}
\toprule
Family & Feature type & What one execution records & Mapped to \\
\midrule
\rowcolor{rowstripe}          & line            & the line was executed                        & own line \\
\rowcolor{rowstripe}          & branch          & the branch took this direction               & construct lines \\
\rowcolor{rowstripe}          & function        & the function was entered                     & construct lines \\
\rowcolor{rowstripe}\multirow{-4}{*}{coverage} & loop & the loop ran zero, one, or many times  & construct lines \\
data flow & definition-use & a value defined here was used there & both lines \\
\rowcolor{rowstripe}          & scalar pair     & a relation between two scalars held          & own line \\
\rowcolor{rowstripe}          & variable        & a variable satisfied a value predicate       & own line \\
\rowcolor{rowstripe}          & return          & a return value satisfied a predicate         & own line \\
\rowcolor{rowstripe}          & none            & a value was \texttt{None}                    & own line \\
\rowcolor{rowstripe}          & length          & a length fell in a regime                    & own line \\
\rowcolor{rowstripe}          & empty string    & a string was empty                           & own line \\
\rowcolor{rowstripe}          & ascii string    & a string was ASCII                           & own line \\
\rowcolor{rowstripe}          & digit string    & a string held digits                         & own line \\
\rowcolor{rowstripe}          & special string  & a string held special characters             & own line \\
\rowcolor{rowstripe}\multirow{-10}{*}{value} & empty bytes & a byte string was empty           & own line \\
condition & condition & a compound condition evaluated true or false & own line \\
\rowcolor{rowstripe}exception & function error  & the function exited through an exception     & construct lines \\
\bottomrule
\end{tabularx}
\end{table}

\subsection{Training the model}

For each subject, we train a \SCIKITLEARN{} \citep{pedregosa2011scikit} random forest of 200 trees with Gini splits on all labeled vectors that remain after constant features are removed.
We do not tune hyperparameters per subject.
All other parameters keep their defaults, except balanced class weights, which account for the imbalance between passing and failing runs when a subject has few failing tests.
Random operations use the master seed $42$ or seeds derived from it deterministically.
We use a random forest because execution features are sparse, correlated, and may be discriminative only in combination.
A forest captures such interactions and yields importance scores that map back to source locations.

The forest serves as a per-subject feature \emph{attribution} device, not as a predictor of unseen data.
It extracts failure-associated importances within a faulty revision.
These are read as per-line weights that modulate an unchanged SFL formula.
The classical overfitting problem to classify unseen executions is not a validity concern since the model is not used as a classifier of unseen data.
Hence, we train the forest per subject without splitting into training and test data.
Moreover, established SFL formulas are transductive and compute suspiciousness from the same pass and fail spectra so that a held-out split would evaluate the augmentation on a different footing than its baseline.
Bagging and per-split feature subsampling reduce variance, and the permutation control (\Cref{sec:negcontrol}) would expose any gain that does not depend on which features the forest selects.

\subsection{Feature importance}\label{sec:mapping}

We extract the \SCIKITLEARN{} Gini importances and map features to lines by their semantics, strictly following the mapping of \EFDD{} \citep{smytzek2026efdd}.
Line features map to their own line, and definition-use pairs map to both the definition and the use line.
Branch, loop, and function features map to all lines of the corresponding construct.
Value predicates map to the line at which the value is observed, which for a variable is the line that assigns it.
These rules were settled before execution and applied strictly for all subjects and formulas.
A feature mapping to multiple lines assigns the same importance to each of them before aggregation.
Raw importances are min--max normalized to $[\varepsilon,1]$ with $\varepsilon=10^{-3}$ as \[ w_f=\varepsilon + (1-\varepsilon)\, \frac{I(f)-I_{\min}}{I_{\max}-I_{\min}} , \] with $I_{\min}=0$ and $I_{\max}=1$ if all importances are equal.
If multiple different features map to the same line, this line carries their maximum normalized weight.
This mechanism keeps many weak features from inflating the weight of a line while allowing a strong failure-indicative feature to affect the ranking.
Executed lines without a mapped feature receive the lower bound $\varepsilon$.
An alternative pre-registered mapping assigns definition-use and scalar-relation importance solely to the line of the usage.
Because scalar-pair features already carry a single source location in \SFLKIT{}, this alternative differs from the primary mapping for definition-use pairs alone, which the protocol did not foresee.

\subsection{Score boosting}

We first compute $\mathrm{susp}_w(l)=\mathrm{susp}'(l)\times w(l)$, where $\mathrm{susp}'(l)$ is the original SFL suspiciousness score and $w(l)$ is the weight derived from the feature importance.
Both terms of the combination are normalized within each subject and formula, such that the scale of an SFL formula does not drive the result.
The primary ranking then uses \[ \mathrm{susp}(l)=(1-\lambda)\,\mathrm{norm}\!\left(\mathrm{susp}'(l)\right) +\lambda\,\mathrm{norm}\!\left(\mathrm{susp}_w(l)\right). \]
The parameter $\lambda$ is registered for the experiments to be $0.35$, and \Cref{sec:constants} explains this and the other constants.
With $1-\lambda=0.65$, the baseline SFL formula remains the dominant term, while $\lambda$ is still large enough for feature evidence to reorder lines where the signals agree.
A smaller value would make the feature signal negligible, and a larger one would dominate the underlying formula.
The pre-registered sensitivity analysis recomputes the primary metrics for $\lambda\in\{0.15,0.25,0.35,0.45,0.55\}$.
It is exploratory and cannot redefine the primary conclusion.

\subsection{Example}\label{sec:example}

\Cref{tab:example} shows the mechanism on \texttt{middle\_1}, the smallest subject of the benchmark. 
The fault in this subject is the assignment in Line~7 of a median-of-three routine.
\TARANTULA{} gives the six lines that the failing run executes four distinct suspiciousness scores.
Among them, lines~2, 3, and 13 share the lowest score because every test executes all three.
The execution-feature weight does differentiate them, since the condition and scalar-pair features observed in the faulty branch occur exclusively in the failing run.
The weighting maintains the order at the top and widens the margin between the fault and its closest competitor from $0.833$ against $0.714$ to $1.000$ against $0.621$.
It also breaks the tie further down in the ranking.
This is the effect the study measures, since the weighting is designed to sharpen an existing ranking rather than to construct a new one.

\begin{table}[t]
\caption{Execution-feature weighting on \texttt{middle\_1} with \TARANTULA{}.
$e_f$ and $e_p$ count the failing and passing executions that observe a line, and $w(l)$ is the normalized execution-feature weight.
The faulty line is set in bold.}\label{tab:example}
\centering
\begingroup\footnotesize\setlength{\tabcolsep}{4pt}
\begin{tabularx}{\linewidth}{rRRRRR}
\toprule
Line & $e_f$ & $e_p$ & $w(l)$ & Tarantula & Augmented \\
\midrule
\rowcolor{rowstripe}\textbf{7} & \textbf{1} & \textbf{1} & \textbf{1.000} & \textbf{0.833} & \textbf{1.000} \\
6 & 1 & 2 & 0.213 & 0.714 & 0.621 \\
\rowcolor{rowstripe}4 & 1 & 3 & 0.213 & 0.625 & 0.544 \\
3 & 1 & 5 & 0.213 & 0.500 & 0.435 \\
\rowcolor{rowstripe}13 & 1 & 5 & 0.106 & 0.500 & 0.413 \\
2 & 1 & 5 & 0.073 & 0.500 & 0.405 \\
\rowcolor{rowstripe}5 & 0 & 1 & 0.213 & 0.000 & 0.000 \\
9 & 0 & 2 & 0.106 & 0.000 & 0.000 \\
\rowcolor{rowstripe}10 & 0 & 1 & 0.000 & 0.000 & 0.000 \\
11 & 0 & 1 & 0.020 & 0.000 & 0.000 \\
\bottomrule
\end{tabularx}
\endgroup
\end{table}

\subsection{Hybrid and adaptive variants}\label{sec:variants}

Beyond the primary weighting approach, we report a small exploratory family of variants.
We explore a \emph{hybrid} variant that combines the available execution signals, a \emph{fail-first} variant that prioritizes discriminative failing evidence, and an \emph{adaptive} variant that reweights samples under class imbalance or sparse failing tests.
The variants reuse the benchmark, mapping and metrics of the primary method and serve only as interpretation aids.
They are corrected for multiplicity separately and cannot substitute for the primary result.
The protocol describes the three variants only qualitatively.
We instantiate them as (i)~a weighted fusion of thirteen per-feature signals computed from the same labeled feature matrix, (ii)~the same fusion with failing evidence up-weighted and passing exposure penalized, and (iii)~a random forest with per-sample weights that scale with the imbalance between passing and failing runs.

\section{Evaluation Procedure}\label{sec:analysis}

The study ran in four phases.
Phases~1 and~2 produce the rankings for \RQ{1}, \RQ{2}, and \RQ{4}; Phase~3 constructs the ground-truth sets for \RQ{3}; and Phase~4 computes the metrics and analyses for all research questions.

\subsection{Evaluation Pipeline}\label{sec:execution}

\paragraph{Phase 1: feature extraction and model training.}
For each \TESTSFOURPY{} subject, we run all relevant tests under \EFDD{} instrumentation with a 120-second timeout per test and convert the traces into labeled feature vectors.
We verify the binary or ternary encoding, train one random forest per subject, and store the fitted importances and feature vectors, so that every ranking can be reproduced without re-running instrumentation.

\paragraph{Phase 2: feature weighting and source-line ranking.}
We extract the importances for features, map these feature-importances to source lines, aggregate them by their maximum, normalize the weights to $[\varepsilon,1]$, and verify that every executable line receives a weight.
For each subject we compute the \TARANTULA{}, \OCHIAI{}, \DSTAR{}, \NAISHTWO{}, and \GPTHIRTEEN{} scores as the baseline.
Then we compute feature-augmented rankings for the same formulas.
Every sampled outcome (top-$k$, impact-found@$k$, and the function-level outcomes) is averaged over $10{,}000$ randomized tie-breaking replicates seeded from the master seed~$42$, so that arbitrary source order cannot decide between equally suspicious lines.

\paragraph{Phase 3: ground-truth construction.}
For each subject, we construct the reference patch-hunk sets from the benchmark's patch files.
Additionally, we construct the test-impact sets with the line-observation rule of \Cref{sec:definitions}.
The two are recorded separately, so the analysis can report whether a method succeeds only on the reference edit, only on broader impact locations, or on both.

\paragraph{Phase 4: metric calculation.}
We calculate the metrics of \Cref{sec:outcomes} and store them by subject, formula, method, and metric, together with the fitted importances and feature vectors.

\subsection{Statistical analysis}\label{sec:stats}

The registered directional expectations are positive reference-patch top-$k$ gains for \RQ{1}, no systematic degradation in line- or function-level effort for \RQ{2}, at least comparable test-impact coverage for \RQ{3}, and a positive direction across most formulas for \RQ{4}.

The primary inference for every outcome is a single confounder-adjusted mixed-effects model fitted over all subjects, formulas, and paired method conditions.
Its fixed effects are the method indicator (augmented against baseline, the effect of interest), the SFL formula as a factor, their interaction, and the confounders of \Cref{sec:confounders} as covariates.
Random intercepts for project and for subject within project, together with a per-project random slope for the method effect, estimate the heterogeneity of the effect directly instead of averaging it away.
Continuous outcomes (\EXAM{}, wasted effort, and function-level effort) use a Gaussian model.

For the per-subject top-$k$ and impact-found@$k$ outcomes, the protocol specifies a binomial logit.
Because these outcomes are averaged over tie-breaking replicates, however, \num{fractionalshare}\% of the per-subject values lie strictly between $0$ and $1$.
The mixed generalized-linear implementation available to us \citep{seabold2010statsmodels} accepts only binary responses.
We therefore use the Gaussian mixed model, read as a linear-probability model, for the superiority rule.
A fractional-response logit \citep{papke1996fractional}, estimated by generalized estimating equations \citep{liang1986gee}, checks that it agrees in sign.
Where the full random-effects structure does not converge, we refit with the project intercept and the per-project method slope only and flag the outcome.
The protocol did not specify this fallback.

We report the method coefficient with its 95\% confidence interval, standardize it for the superiority rule, and read $H_{\mathrm{ROB}}$ from the interaction.
The confounder coefficients and the random-slope variance characterize where and why the effect varies.
The method coefficients are corrected with the Benjamini--Hochberg procedure \citep{benjamini1995fdr} within the reference-patch top-$k$, effort, and test-impact families.
Besides them, we report improve, tie, and degrade counts per research question.

To corroborate the model under minimal distributional assumptions, we re-test each contrast with distribution-free paired procedures.
For the top-$k$ and impact-found@$k$ proportions, these are paired permutation tests with paired mean differences and 95\% bootstrap confidence intervals \citep{efron1992bootstrap}.
For \EXAM{}, wasted effort, function-weighted \EXAM{}, faulty-function mass, practical effort, and minimum impact rank, they are Wilcoxon signed-rank tests \citep{wilcoxon1992ranksum} with rank-biserial correlations \citep{kerby2014rankbiserial}.
Patch-hunk and test-impact sets are analyzed separately.

\subsection{Controls}\label{sec:controls}

\paragraph{Data quality.}
Before interpreting results, we verify that all collected artifacts, i.e., the labeled feature vectors, their encodings, the line weights, the patch-hunk, and test-impact records, can be evaluated.
If a check fails, we record the affected subject and either repair the processing step or exclude the subject from the paired analysis with an explicit reason.
Implementation failures thus cannot be read as localization performance.
We also retrain ten randomly selected subjects with the same seed to verify identical metrics, and we inspect the paired differences of any subject that does not match.

\paragraph{Positive control.}
The positive control checks that the pipeline propagates evidence about a known fault to the ranking.
The protocol specifies this as an auxiliary feature that separates passing from failing runs perfectly.
Added to the forest's input, such a feature absorbs importance mass, deflates the real features through the normalization, and often receives no importance of its own.
Hence, the feature did not measure what it was meant to measure.
We instead implement the positive control as an oracle weighting, i.e., the learned weights with every patch reference line increased to the largest possible weight.
This construction dominates the learned weighting on every subject and expresses perfect knowledge through the registered mapping and blend.

\paragraph{Negative control.}\label{sec:negcontrol}
The negative control is the direct guard against spurious importance scores.
This control replaces the learned importances with random scores and checks that an improvement is not simply a result of modifying the suspiciousness scores.
Running the registered negative control revealed that it is not a pure noise control.
A line's weight is the \emph{maximum} over the features mapped to it, and the maximum of $k$ independent uniform draws increases with $k$.
The random weight is largely a monotone transform of how many features map to a line.
Across subjects, its per-line weight correlates with the mapped-feature count at a median Spearman $\rho$ of \num{negdensityrho}, against \num{rfdensityrho} for the learned weights.
Feature multiplicity is a structural property of the code, so a gain of the random control over the baseline would not show that perturbing the scores is enough on its own.

Therefore, we decided to add a control that does not alter this property.
It keeps the learned importance \emph{distribution} and permutes it across features.
Variety and the shape of the distribution stay unchanged, and only the relationship between a feature and its importance is destroyed.

Overall, these controls answer the following questions.
The registered negative control asks whether a line's feature density carries signal.
The permutation control asks whether the learned association carries signal beyond that.
Finally, the positive control asks whether the pipeline can propagate a known cause at all.
\Cref{tab:methods} reports all three.
The permutation control is an addition that was not registered, and the registered negative control is reported unchanged.

\paragraph{Controls on code.}
Either verdict is informative only if the analysis detects an effect when there is one and stays silent when there is none.
We added a software-level control that is independent of the study data.
It generates a synthetic study of the registered size (310 subjects, 20 projects, five formulas, paired conditions) in which the augmented method is better by a known margin and checks that the decision rule fires.
It then generates the same study without an effect and checks that the rule stays silent.
Both checks pass.
The implementation's power calculation reproduces the registered detectable effect of $d_z\approx0.16$.

\subsection{Exploratory analyses}\label{sec:diagnostics}

To interpret the results, we compute diagnostics from the artifacts without manual relabeling of \TESTSFOURPY{}.
These diagnostics include feature-family contributions, project-level variation, constant-feature fractions, importance concentration, entropy, skewness, failing-vector counts, mapping-conflict rates, and the correlations between outcomes.
The registered sensitivity checks vary the weight floor $\varepsilon\in\{10^{-4},10^{-3},10^{-2}\}$, the blend parameter $\lambda\in\{0.15,0.25,0.35,0.45,0.55\}$, and the aggregation operator (mean and sum instead of maximum).
Sum aggregation can push a line's weight above one, so for that condition only we min--max renormalize the aggregated values before applying the floor.
Where at least three passing and three failing vectors exist, three seeded and three bootstrap refits per subject measure the stability of the importances.
The protocol leaves the number of refits open, and six keep the diagnostic affordable for subjects with hundreds of thousands of features.

Within the registered evaluation, several choices go beyond or deviate from the protocol, namely the permutation control (\Cref{sec:negcontrol}), the oracle positive control (\Cref{sec:controls}), the Gaussian model and its fallback structure (\Cref{sec:stats}), the deterministic \EXAM{} and wasted effort and the paired effect size $d_z$ in the decision rule (\Cref{sec:decision}), and deciding $H_{\mathrm{ROB}}$ on the superiority rule (\Cref{sec:rq4}).
Each is marked where it appears, and the registered $d$ gives the same verdict for every registered method.
The ranker that learns across projects (\Cref{sec:ltr}) was designed after the registered results were known and is reported separately.

\subsection{Missing data}

A subject is considered for the paired analysis if at least one passing and one failing feature vector could be collected while running the tests.
It is excluded only if it cannot be executed, if feature extraction fails, or if one outcome is absent.
We set no threshold on the number of failing tests, except that there must exist one.
Such a threshold would select against sparse but realistic subjects.

\section{Results}\label{sec:results}

\subsection{Feasibility}

The pipeline completed on all \num{nsubjects} subjects from \num{nprojects} projects (\Cref{tab:subjects}).
Every subject produced at least one passing and one failing execution feature vector, so the analysis covers all registered subjects without any exclusions.
Hence, our evaluation reports every registered outcome.

The cost is concentrated in a single phase.
The median subject took \num{medianruntime}\,s end-to-end, and the whole study consumed \num{totalruntime} core-hours.
Converting traces into labeled feature vectors dominates this total.
The three steps that the method itself contributes, fitting the forest, mapping features to lines, and blending the weights, account for a negligible share.
Execution-feature weighting is cheap to add to an existing spectrum-based pipeline and expensive only in the instrumentation that any execution-feature approach must pay for.

\subsection{\texorpdfstring{\RQ{1}: patch-line accuracy}{\RQ{1}: patch-line accuracy}}\label{sec:rq1}

\begin{figure}[t]
\centering
\includegraphics[width=\linewidth]{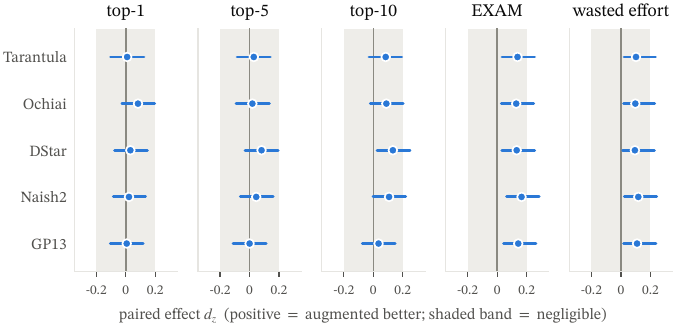}
\caption{Confounder-adjusted method effect of execution-feature weighting per formula and primary outcome.
Points are paired effects $d_z$ with 95\% bootstrap confidence intervals, oriented so that positive always means that the augmented method is better.
The shaded band marks the negligible region $|d_z|<0.2$.}\label{fig:forest}
\end{figure}

Execution-feature weighting barely moves the ranking (\Cref{fig:forest} and \Cref{tab:primary}).
Averaged over the five formulas, top-1 accuracy goes from \num{basepatchtop1} to \num{augpatchtop1}.
Every paired effect sits inside the negligible region $|d_z|<0.2$.

\Cref{tab:superiority} applies the decision rule to each formula individually.
The paired effect reaches at most \num{rfmaxabsdz} on any primary outcome, so no formula improves a single one by $|d_z|\ge0.2$, and none degrades one.
The registered $d$ is smaller still, at most \num{rfmaxabsd}, and gives the same verdict.

For individual subjects, the picture is mixed.
On top-1, the fault moves down the ranking on more subjects than it moves up, and the small gain in the mean comes from fewer but larger promotions.
Read one subject at a time (\Cref{tab:per-subject}), the method is superior on \num{pssuperior}\% of subject-formula pairs and inferior on \num{psinferior}\%.
On an individual debugging task, the weighting is thus about as likely to hurt as to help.

Of the \num{nmodels} mixed models fitted for the primary method, \num{nconverged} converged with the full random-effects structure.
The others, among them the model for top-5, fall back to the simpler structure of \Cref{sec:stats} and are flagged.
The fractional-response GEE agrees with the Gaussian model in sign on every top-$k$ outcome, and neither model finds a significant method effect on any of them.

\begin{finding}[\RQ{1}: Execution-feature weighting does not improve accuracy]
No primary outcome reaches $|d_z|\ge0.2$ for any formula (largest $|d_z|=\num{rfmaxabsdz}$), the method is superior for \num{npairedsuperior} of 5 formulas, and $H_{\mathrm{ACC}}$ is not supported.
The registered $d$ gives the same verdict, and on an individual debugging task the weighting is about as likely to hurt as to help.
\end{finding}

\begin{table}[t]
\caption{Primary reference-patch outcomes per formula.
\emph{Baseline} and \emph{Augmented} are means over subjects.
$\Delta$ is the confounder-adjusted coefficient of the mixed model, from which $d$, the confidence interval, and $q$ are derived, so it can differ from the difference of the two means.
$d_z$ is computed from the per-subject differences and determines the rule.}\label{tab:primary}
\centering
\begingroup\scriptsize\setlength{\tabcolsep}{3.2pt}
\begin{tabularx}{\linewidth}{lLrrrrrrr}
\toprule
Formula & Outcome & Baseline & Augmented & $\Delta$ & $d$ & $d_z$ & 95\% CI & $q$ \\
\midrule
\rowcolor{rowstripe} & top-1 & 0.167 & 0.170 & 0.002 & 0.01 & 0.01 & [-0.034, 0.039] & 0.944 \\
\rowcolor{rowstripe} & top-5 & 0.413 & 0.422 & 0.011 & 0.03 & 0.03 & [-0.040, 0.063] & 0.944 \\
\rowcolor{rowstripe} & top-10 & 0.523 & 0.547 & 0.027 & 0.06 & 0.08 & [-0.017, 0.072] & 0.814 \\
\rowcolor{rowstripe} & EXAM & 0.182 & 0.173 & -0.008 & -0.03 & -0.14 & [-0.035, 0.019] & 0.838 \\
\rowcolor{rowstripe}\multirow{-5}{*}{Tarantula} & WE & 625 & 591 & -21 & -0.01 & -0.10 & [-202, 160] & 0.838 \\
 & top-1 & 0.160 & 0.181 & 0.021 & 0.07 & 0.08 & [-0.016, 0.057] & 0.814 \\
 & top-5 & 0.406 & 0.412 & 0.008 & 0.02 & 0.02 & [-0.044, 0.060] & 0.944 \\
 & top-10 & 0.514 & 0.538 & 0.027 & 0.06 & 0.09 & [-0.017, 0.071] & 0.814 \\
 & EXAM & 0.194 & 0.187 & -0.007 & -0.03 & -0.13 & [-0.034, 0.020] & 0.838 \\
\multirow{-5}{*}{Ochiai} & WE & 633 & 601 & -19 & -0.01 & -0.10 & [-200, 161] & 0.838 \\
\rowcolor{rowstripe} & top-1 & 0.110 & 0.116 & 0.006 & 0.02 & 0.03 & [-0.031, 0.043] & 0.944 \\
\rowcolor{rowstripe} & top-5 & 0.254 & 0.272 & 0.021 & 0.05 & 0.08 & [-0.031, 0.073] & 0.944 \\
\rowcolor{rowstripe} & top-10 & 0.332 & 0.362 & 0.034 & 0.08 & 0.13 & [-0.011, 0.078] & 0.814 \\
\rowcolor{rowstripe} & EXAM & 0.386 & 0.379 & -0.008 & -0.02 & -0.13 & [-0.034, 0.019] & 0.838 \\
\rowcolor{rowstripe}\multirow{-5}{*}{DStar} & WE & 1,207 & 1,174 & -19 & -0.01 & -0.09 & [-200, 161] & 0.838 \\
 & top-1 & 0.164 & 0.171 & 0.006 & 0.02 & 0.02 & [-0.031, 0.043] & 0.944 \\
 & top-5 & 0.400 & 0.414 & 0.016 & 0.04 & 0.05 & [-0.035, 0.068] & 0.944 \\
 & top-10 & 0.498 & 0.528 & 0.034 & 0.07 & 0.11 & [-0.010, 0.078] & 0.814 \\
 & EXAM & 0.212 & 0.198 & -0.014 & -0.05 & -0.17 & [-0.041, 0.013] & 0.828 \\
\multirow{-5}{*}{Naish2} & WE & 651 & 600 & -38 & -0.03 & -0.12 & [-219, 143] & 0.838 \\
\rowcolor{rowstripe} & top-1 & 0.164 & 0.166 & 0.001 & 0.00 & 0.01 & [-0.035, 0.038] & 0.944 \\
\rowcolor{rowstripe} & top-5 & 0.400 & 0.400 & 0.003 & 0.01 & 0.00 & [-0.049, 0.054] & 0.944 \\
\rowcolor{rowstripe} & top-10 & 0.498 & 0.509 & 0.015 & 0.03 & 0.04 & [-0.029, 0.059] & 0.944 \\
\rowcolor{rowstripe} & EXAM & 0.212 & 0.200 & -0.011 & -0.04 & -0.14 & [-0.038, 0.015] & 0.828 \\
\rowcolor{rowstripe}\multirow{-5}{*}{GP13} & WE & 685 & 637 & -35 & -0.02 & -0.11 & [-215, 146] & 0.838 \\
\bottomrule
\end{tabularx}
\endgroup
\end{table}

\begin{table}[t]
\caption{The superiority rule per formula, applied with the paired effect $d_z$.
An outcome counts as improved only if $|d_z|\ge0.2$ in the beneficial direction.
The registered $d$ gives the same verdict.}\label{tab:superiority}
\centering
\begingroup\scriptsize\setlength{\tabcolsep}{3.2pt}
\begin{tabularx}{\linewidth}{Lccccccl}
\toprule
Formula & top-1 & top-5 & top-10 & EXAM & WE & improved & superior \\
\midrule
\rowcolor{rowstripe}Tarantula & $\approx$\,0.01 & $\approx$\,0.03 & $\approx$\,0.08 & $\approx$\,0.14 & $\approx$\,0.10 & 0/5 & no \\
Ochiai & $\approx$\,0.08 & $\approx$\,0.02 & $\approx$\,0.09 & $\approx$\,0.13 & $\approx$\,0.10 & 0/5 & no \\
\rowcolor{rowstripe}DStar & $\approx$\,0.03 & $\approx$\,0.08 & $\approx$\,0.13 & $\approx$\,0.13 & $\approx$\,0.09 & 0/5 & no \\
Naish2 & $\approx$\,0.02 & $\approx$\,0.05 & $\approx$\,0.11 & $\approx$\,0.17 & $\approx$\,0.12 & 0/5 & no \\
\rowcolor{rowstripe}GP13 & $\approx$\,0.01 & $\approx$\,0.00 & $\approx$\,0.04 & $\approx$\,0.14 & $\approx$\,0.11 & 0/5 & no \\
\bottomrule
\end{tabularx}
\endgroup
\end{table}

\begin{table}[t]
\caption{The superiority rule for one subject at a time over all subject-formula pairs.
The pooled rule asks whether the method wins on average, and this table asks how often it wins on an individual debugging task.}\label{tab:per-subject}
\centering
\begingroup\footnotesize\setlength{\tabcolsep}{4pt}
\begin{tabularx}{\linewidth}{lR}
\toprule
Per-subject outcome & Value \\
\midrule
\rowcolor{rowstripe}subject/formula pairs & 1550 \\
mean primary outcomes improved & 1.09 of 5 \\
\rowcolor{rowstripe}mean primary outcomes degraded & 1.35 of 5 \\
superior ($\ge$4 better, none worse) & 6.9\% \\
\rowcolor{rowstripe}inferior ($\ge$4 worse, none better) & 9.7\% \\
more outcomes better than worse & 29.9\% \\
\rowcolor{rowstripe}more outcomes worse than better & 46.3\% \\
identical on all five & 17.1\% \\
\bottomrule
\end{tabularx}
\endgroup
\end{table}

\subsection{\texorpdfstring{\RQ{2}: developer effort}{\RQ{2}: developer effort}}\label{sec:rq2}

Averaged over the five formulas, \EXAM{}, where lower is better, goes from \num{basepatchexamavg} to \num{augpatchexamavg}.
\EXAM{} gets worse on \num{rfdegradeexam} subjects and better on \num{rfimproveexam}, and the Wilcoxon tests find this shift toward a worse \EXAM{} for \num{wilexamnreject} of the five formulas.
The mean still improves slightly, because the improvements are larger.
Function-level effort is just as small, with function-level \EXAM{} at $d=\num{dfunctionexam}$.

The function-level mass outcomes need care, because their pooled effect averages over a disagreement between formulas.
The joint interaction between weighting and formula is significant ($p\,\num{faminteraction}$).
Four of the five formulas improve by $d=\num{othersdmin}$ to $\num{othersdmax}$, while \num{negformula} degrades by $d=\num{naishd}$, and \Cref{sec:ceiling} takes up what drives the disagreement.

\begin{finding}[\RQ{2}: Execution-feature weighting does not reduce effort]
Neither \EXAM{} nor WE reaches $|d_z|\ge0.2$ for any formula, and function-level \EXAM{} changes just as little, so effort is maintained on average.
On individual subjects, however, \EXAM{} gets worse on \num{rfdegradeexam} and better on \num{rfimproveexam}, a shift that the Wilcoxon tests find for \num{wilexamnreject} of 5 formulas.
\end{finding}

\subsection{\texorpdfstring{\RQ{3}: broader ground truth}{\RQ{3}: broader ground truth}}\label{sec:rq3}

The reference patch is a narrow target.
A developer who lands on the enclosing function, or on a line that the failing test demonstrably depends on, has usually made real progress.
\RQ{3} therefore repeats the comparison against the reference patch hunks and the dynamically discovered test-impact sets.
Beside them, \Cref{sec:rq2} reports the function-level effort as a third relaxation of the reference patch.

Hunk-level outcomes, which credit any line of a reference hunk, move with the patch-line outcomes and by a similarly negligible amount, with $d=\num{dhunktop1}$ for hunk@1 and $d=\num{dhunktop10}$ for hunk@10.
Relaxing the ground truth from the line to the hunk or the enclosing function does not uncover an effect that the reference patch hides.

Test-impact outcomes (\Cref{tab:impact}) move the other way.
Impact@1 falls with $d=\num{dimpacttop1}$, the largest pooled effect of the primary method on any outcome, while impact@5 and impact@10 are essentially unchanged.
Most subject-formula pairs are tied on these outcomes, because the baseline already places an impact location first for the large majority of subjects.
Where the weighting changes the first position, it far more often displaces an impact location than puts one there.
The patch line and the impact locations compete for the same top rank, so promoting the one can demote the other.

\begin{finding}[\RQ{3}: A broader ground truth reveals no hidden effect]
Crediting any line of a reference hunk or the enclosing function uncovers no improvement that the patch lines hide.
On test-impact locations the weighting costs the first position, with impact@1 at $d=\num{dimpacttop1}$, the largest pooled effect of the method on any outcome.
\end{finding}

\begin{table}[t]
\caption{Coverage of dynamically discovered test-impact locations.}\label{tab:impact}
\centering
\begingroup\scriptsize\setlength{\tabcolsep}{3.2pt}
\begin{tabularx}{\linewidth}{Lrrrrrrrr}
\toprule
Outcome & Baseline & Augmented & $\Delta$ & $d$ & $q$ & Improve & Tie & Degrade \\
\midrule
\rowcolor{rowstripe}impact@1 & 0.866 & 0.820 & -0.046 & -0.16 & 0.007 & 5 & 1,468 & 77 \\
impact@5 & 0.870 & 0.863 & -0.006 & -0.04 & 1.000 & 13 & 1,514 & 23 \\
\rowcolor{rowstripe}impact@10 & 0.870 & 0.874 & 0.004 & 0.00 & 1.000 & 19 & 1,517 & 14 \\
min.\ impact rank & 269 & 245 & -23 & -0.04 & 1.000 & 115 & 1,282 & 153 \\
\rowcolor{rowstripe}impact locations & 429 & 429 & 0 & 0.00 & 1.000 & 0 & 1,550 & 0 \\
\bottomrule
\end{tabularx}
\endgroup
\end{table}

\subsection{\texorpdfstring{\RQ{4}: robustness and heterogeneity}{\RQ{4}: robustness and heterogeneity}}\label{sec:rq4}

The method is superior for none of the five formulas (\Cref{tab:superiority}), so the condition that the protocol sets for $H_{\mathrm{ROB}}$ is not met.
Only the absence of an effect holds across formulas.
On top-1, the effect ranges only from $d=\num{worstformulad}$ for \num{worstformula} to $d=\num{bestformulad}$ for \num{bestformula}, and the method$\times$formula interaction is not significant ($p\,\num{topinteraction}$).

The protocol asks that superiority for all formulas be confirmed by the interaction and a sign test.
With no formula superior, there is nothing to confirm, and the sign test gives $p=\num{signtestp}$.
A two-sided sign test over five formulas cannot reject at $\alpha=0.05$ whatever the data, which the protocol did not anticipate.
We therefore decide $H_{\mathrm{ROB}}$ on the superiority rule, as for $H_{\mathrm{ACC}}$.

\begin{figure}[t]
\centering
\includegraphics[width=\linewidth]{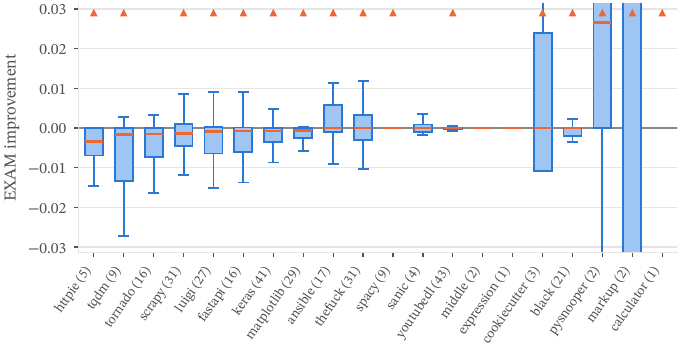}
\caption{Per-project distribution of the paired change in \EXAM{}.
Positive values are improvements, and the boxes span the interquartile range over all subject-formula pairs of a project, and the number of subjects follows each project name.
A few very small projects produce extreme changes, so the axis is clipped to the bulk of the distribution.
A triangle marks a project whose spread runs past it.}\label{fig:projects}
\end{figure}

\Cref{fig:projects} shows the paired change in \EXAM{} by project.
On top-1, the mean paired change is positive in \num{npositiveprojects} of the \num{nprojects} projects, about what a null effect with project-level noise would produce.
We also regressed the paired change in each primary outcome on eight subject properties that ought to predict when learned weights help.
They are the share of lines whose weight rises above the floor, the fraction of constant features, importance concentration, entropy, the share of features that map to several lines, the number of failing tests, program size, and the weak-signal flag.
Of the \num{nmoderatortests} coefficients fitted across outcomes and properties, \num{nmoderatorsig} reach nominal significance against \num{nmoderatorexpected} expected by chance at $\alpha=0.05$, and \num{nmoderatorholm} remain after a Holm correction.
None of the properties identify a subpopulation on which the weighting helps.
This includes the number of failing tests.
Within the evaluated benchmark, subjects with more failing tests do not benefit measurably more.

\Cref{tab:methods} extends the comparison to the alternative mapping and the exploratory ranking variants.
The alternative mapping behaves like the primary one (top-1 \num{alttopone}).
The adaptive variant, which reweights the forest's samples under class imbalance, also behaves like the primary one (top-1 \num{adaptivetopone}).
The two fusion variants, hybrid and fail-first, stay at the baseline (top-1 \num{hybridtopone} and \num{failfirsttopone}).
None of them is superior for any formula, under $d$ or under $d_z$.
The registered exploratory family offers no alternative aggregation of the same feature matrix that works where the forest does not.

\begin{finding}[\RQ{4}: No formula benefits]
The method is superior for none of the five formulas, so $H_{\mathrm{ROB}}$ is not supported.
The effect on top-1 lies between $d=\num{worstformulad}$ (\num{worstformula}) and $d=\num{bestformulad}$ (\num{bestformula}); no project or subject property marks a subpopulation that benefits; and no registered variant does better.
\end{finding}

\begin{table}[t]
\caption{The registered method beside its exploratory variants and the outcome-neutral controls, as means over subjects and the five formulas.
The controls replace the learned importances with uniform random ones or with the learned ones permuted across features, or give the fault lines the largest weight that the pipeline admits.}\label{tab:methods}
\centering
\begingroup
\begin{tabularx}{\linewidth}{llRRRRR}
\toprule
 & Condition & top-1 & top-5 & top-10 & EXAM & WE \\
\midrule
\rowcolor{rowstripe} & baseline SFL & 0.153 & 0.375 & 0.473 & 0.237 & 760 \\
\rowcolor{rowstripe}\multirow{-2}{*}{Method} & RF weighting & 0.161 & 0.384 & 0.497 & 0.228 & 721 \\
 & alternative mapping & 0.163 & 0.385 & 0.493 & 0.227 & 721 \\
 & hybrid & 0.151 & 0.373 & 0.472 & 0.237 & 760 \\
 & fail-first & 0.152 & 0.371 & 0.471 & 0.237 & 760 \\
\multirow{-4}{*}{Variants} & adaptive & 0.149 & 0.380 & 0.496 & 0.229 & 722 \\
\rowcolor{rowstripe} & random importances & 0.173 & 0.401 & 0.498 & 0.224 & 662 \\
\rowcolor{rowstripe} & permuted importances & 0.143 & 0.360 & 0.476 & 0.231 & 739 \\
\rowcolor{rowstripe}\multirow{-3}{*}{Controls} & oracle (positive) & 0.494 & 0.724 & 0.779 & 0.140 & 393 \\
\bottomrule
\end{tabularx}
\endgroup
\end{table}

\subsection{Outcome-neutral controls}\label{sec:controlresults}

The controls decide whether the absence of an effect lies in the setup or in the learned importances, and the last group of \Cref{tab:methods} reports them.

The positive control gives the true fault lines the largest weight that the pipeline can assign.
It raises top-1 accuracy from \num{basepatchtop1} to \num{postop1} and is superior for \num{posnpairedsuperior} of 5 formulas.
The mapping and the blend therefore carry a signal when there is one, and an effect of that kind would have been detected easily.
The control also marks the upper bound for any weighting expressed through the registered blend.
The learned weighting captures only \num{headroomshare}\% of the top-1 headroom that this bound leaves.

The registered negative control replaces the importances with uniform random weights.
It does at least as well as the learned weighting.
It reaches top-1 \num{negtop1} against \num{augpatchtop1} for the forest.
Like the forest, it is superior for no formula.
As \Cref{sec:negcontrol} explains, its random weights track how many features map to a line, so its slight advantage reflects densely instrumented lines rather than the failure.

The permutation control permutes the learned importances across features within a subject.
This method preserves the importance distribution and the mapping structure.
It only eliminates the association between a feature and its weight.
It reaches top-1 \num{permtop1}, slightly below both the baseline and the registered method.
These controls bound what the forest learned.
Neither the random nor the permuted importances differ from the baseline by a non-negligible effect, but neither do the learned ones.

\subsection{Where the evidence comes from}

\begin{figure}[t]
\centering
\includegraphics[width=\linewidth]{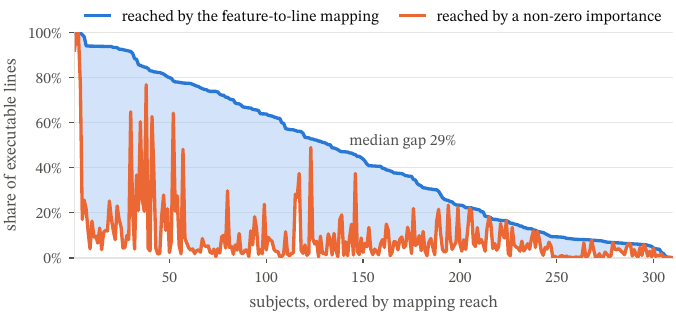}
\caption{How far execution-feature evidence reaches down to individual lines.
The upper curve is the share of executable lines that the feature-to-line mapping reaches at all, and the lower curve is the share that a feature with non-zero random-forest importance reaches.
The mapping reaches the upper curve, so the gap between the curves comes from the model.}\label{fig:bottleneck}
\end{figure}

The median number of failing test executions per subject is \num{medianfailing}. Moreover,
With a single failing run, every feature whose value in that run differs from all passing runs separates the two classes perfectly, and there are typically many such features.
After the \num{medianconstant}\% of features that are constant across the labeled runs are removed, the forest assigns non-zero importance to a median of only \num{mediannonzero} of \num{medianfeatures} features.
Those features map onto \num{medianimportancereach}\% of the executable lines (\Cref{fig:bottleneck} and \Cref{tab:diagnostics}).
Which of the many perfect separators the forest credits is decided by its bootstrap and feature subsampling rather than by the data.
Not all of them are suited for localization, since a feature can separate the runs where the infection shows rather than where the fix applies.
The importances are correspondingly unstable.
Even among the better-supplied subjects with at least three failing and three passing runs, a refit with another seed shares a median of only \num{seededjaccard} of its top-20 features with the original.
In the worked example of \Cref{tab:example}, the forest happens to credit features at the faulty branch, but across the benchmark it does so no more often than random weights do.

Among the feature families, value features carry the largest share of the importance mass, \num{sharevalue}\%.
The primary outcomes are correlated with each other, most strongly top-5 with top-10 and \EXAM{} with wasted effort, so agreement among them is less independent evidence than five outcomes suggest.

Finally, in \num{nounreached} subjects no test executes any reference line.
No line-level ranking of this kind can succeed on them, and they stay in the analysis, because excluding them after seeing the results would bias it.

\begin{finding}[Why: The learned importances, not the pipeline, are the problem]
An oracle expressed through the same mapping and blend is superior for every formula, and random importances do as well as learned ones.
With a single failing run in \num{shareonefailing}\% of subjects, multiple features separate it equally well, and not all of them sit where the fix applies.
\end{finding}

\begin{table}[t]
\caption{Diagnostics over all analyzed subjects.}\label{tab:diagnostics}
\centering
\begingroup\footnotesize\setlength{\tabcolsep}{4pt}
\begin{tabularx}{\linewidth}{lRRRR}
\toprule
Diagnostic & Median & IQR low & IQR high & $n$ \\
\midrule
\rowcolor{rowstripe}execution features & 62,425 & 17,710 & 187,481 & 310 \\
non-constant features & 28,728 & 5,540 & 116,301 & 310 \\
\rowcolor{rowstripe}constant fraction & 0.543 & 0.155 & 0.792 & 310 \\
failing vectors & 1 & 1 & 2 & 310 \\
\rowcolor{rowstripe}passing vectors & 18 & 8 & 50 & 310 \\
multi-location features & 0.032 & 0.020 & 0.054 & 310 \\
\rowcolor{rowstripe}lines reached by the mapping & 0.408 & 0.136 & 0.720 & 310 \\
lines reached by a non-zero importance & 0.046 & 0.020 & 0.104 & 310 \\
\rowcolor{rowstripe}lines above the weight floor & 0.046 & 0.020 & 0.104 & 310 \\
features with non-zero importance & 191 & 142 & 288 & 310 \\
\rowcolor{rowstripe}importance Gini & 0.996 & 0.975 & 0.999 & 310 \\
importance entropy & 0.483 & 0.428 & 0.587 & 310 \\
\rowcolor{rowstripe}importance skewness & 18.00 & 7.85 & 32.91 & 310 \\
seeded refit $\rho$ & 0.204 & 0.082 & 0.408 & 42 \\
\rowcolor{rowstripe}bootstrap refit $\rho$ & 0.376 & 0.197 & 0.528 & 42 \\
seeded top-20 Jaccard & 0.039 & 0.018 & 0.223 & 42 \\
\rowcolor{rowstripe}bootstrap top-20 Jaccard & 0.123 & 0.053 & 0.373 & 42 \\
\bottomrule
\end{tabularx}
\endgroup
\end{table}

\subsection{Sensitivity}

The verdict does not depend on the registered parameter values (\Cref{tab:sensitivity}).
Across the blend interval $\lambda\in[\num{lambdamin},\num{lambdamax}]$, the augmented ranking stays close to the baseline, and it gets slightly \emph{worse} as $\lambda$ grows (top-1 from \num{lambdaattmin} to \num{lambdaattmax}).
The more weight the learned features get, the lower top-$k$ accuracy becomes, while \EXAM{} and WE barely change, so no value of $\lambda$ would have rescued the method.
The weight floor $\varepsilon$ is inert, and the aggregation operator makes no practical difference.
The test-impact threshold changes the size of the impact target but not the direction of the comparison.
At every threshold from 0.5 to 0.8, impact@1 of the augmented ranking stays below the baseline's.

\begin{table}[t]
\caption{Registered sensitivity of the primary outcomes to the parameters that move the ranking.}\label{tab:sensitivity}
\centering
\begingroup\scriptsize\setlength{\tabcolsep}{3.2pt}
\begin{tabularx}{\linewidth}{lLrrrrr}
\toprule
Knob & Value & top-1 & top-5 & top-10 & EXAM & WE \\
\midrule
\rowcolor{rowstripe} & 0.15 & 0.167 & 0.392 & 0.500 & 0.228 & 721 \\
\rowcolor{rowstripe} & 0.25 & 0.166 & 0.393 & 0.500 & 0.228 & 721 \\
\rowcolor{rowstripe} & 0.35 & 0.161 & 0.384 & 0.497 & 0.228 & 721 \\
\rowcolor{rowstripe} & 0.45 & 0.158 & 0.383 & 0.495 & 0.227 & 721 \\
\rowcolor{rowstripe}\multirow{-5}{*}{$\lambda$} & 0.55 & 0.157 & 0.377 & 0.493 & 0.225 & 720 \\
 & 0.0001 & 0.161 & 0.384 & 0.496 & 0.228 & 721 \\
 & 0.001 & 0.161 & 0.384 & 0.497 & 0.228 & 721 \\
\multirow{-3}{*}{$\varepsilon$} & 0.01 & 0.160 & 0.384 & 0.496 & 0.228 & 721 \\
\rowcolor{rowstripe} & max & 0.161 & 0.384 & 0.497 & 0.228 & 721 \\
\rowcolor{rowstripe} & mean & 0.165 & 0.383 & 0.486 & 0.229 & 727 \\
\rowcolor{rowstripe}\multirow{-3}{*}{aggregation} & sum & 0.162 & 0.391 & 0.505 & 0.227 & 720 \\
\bottomrule
\end{tabularx}
\endgroup
\end{table}

\subsection{Learning across projects}\label{sec:ltr}

The registered results locate the failure in the attribution step.
In \num{shareonefailing}\% of subjects, a learner fitted to one subject has a single failing run to learn from.
Multiple features separate that run equally well, and the learner cannot tell which of them sit where the fix applies.
A learner that has seen the faults of other projects is not limited in this way, because it can learn which kind of evidence tends to mark a faulty line.
We therefore turn the tables and let a ranker learn across projects.
This analysis was designed after the registered results were known and does not change the verdict, but it uses the registered subjects, outcomes, and decision rule.

\paragraph{Approach.}
The ranker describes each line by a set of inputs from three families.
\begin{description}
\item[SFL inputs.] They are the scores of the five formulas and the share of failing and passing runs that execute the line.
\item[Feature inputs.] They are three line weights computed from the stored feature matrices.
The first is the forest importance of the registered method.
The second, a feature spectrum, scores each feature value by its \OCHIAI{} coefficient.
The third, a nearest-neighbor contrast, compares each failing run with its most similar passing run and credits only the features in which the two differ \citep{renieris2003nearest}.
All three reach the lines through the registered mapping and maximum aggregation.
\item[Trace inputs.] They use the order in which events happen.
For each failing run, we compute its dynamic backward slice \citep{agrawal1993debugging, korel1988dynamic}, starting from the values that reach the test and from the origin of an exception that reaches it.
We also record how recently each line ran before the failure.
Across all runs, we build an automaton whose states are the executed lines.
We score its transitions, namely control-flow edges, calls, and paths of three lines, by \OCHIAI{} and by nearest-neighbor contrast, as for the features.
\end{description}
Each input also enters as its maximum over the line's function and over its file.
To make subjects of different size comparable, every value becomes a percentile rank within its subject.

The feature vectors do not keep the order of events.
To obtain the trace inputs, we therefore re-ran the subjects' existing tests with ordered tracing, without generating any test.
These traces come from current releases of \SFLKIT{} and \TESTSFOURPY{}, so we match their lines to the study's lines by file and line number.
For every subject, at least \num{ltrtracecoveragemin}\% of the lines match.
Unmatched lines get weight zero, and one subject has no trace inputs at all.

The ranker is a gradient-boosted tree classifier \citep{pedregosa2011scikit} that predicts for each line whether it is a reference line.
We fixed its settings before its first run, i.e., balanced class weights, 300 trees of at most 31 leaves, a learning rate of 0.05, and at least 50 lines per leaf.
We train it \emph{leave one project out}.
To rank the lines of a \texttt{matplotlib} bug, for example, the model is trained only on the bugs of the other 19 projects and has seen no \texttt{matplotlib} bug at all.
Every project is held out in turn.
The percentiles of the inputs are computed within each subject, so they carry no information from other subjects either.

The ranking orders lines by predicted probability and breaks ties by the formula's score.
We compare it with each formula's baseline exactly as the registered method.
Because the ranker replaces the registered blend, the ceiling of \Cref{sec:ceiling} does not apply to it.

Two controls and one ablation show where a gain comes from.
The controls train the same model with random or permuted importances in place of the feature inputs, and the ablation trains it on the SFL inputs alone.
If the full ranker beats all three, its gain comes from what the features say, and not from learning across subjects or from feature multiplicity (\Cref{sec:negcontrol}).

\begin{table}[t]
\caption{A ranker that learns across projects, trained with leave-one-project-out.
Each row adds inputs to the SFL scores, and the controls use random or permuted importances in place of the feature weightings.
The rows \emph{+ feature vectors} and \emph{+ trace signals} add the feature inputs and the trace inputs.
Outcomes are means over subjects and the five formulas.
The last two columns count the formulas for which the four-of-five rule holds under the registered $d$ and under the paired $d_z$, which decides it.}\label{tab:ltr}
\centering
\begingroup\footnotesize\setlength{\tabcolsep}{4pt}
\begin{tabularx}{\linewidth}{lRRRRRCC}
\toprule
Method & top-1 & top-5 & top-10 & EXAM & WE & sup.\ ($d$) & sup.\ ($d_z$) \\
\midrule
\rowcolor{rowstripe}baseline & 0.153 & 0.375 & 0.473 & 0.237 & 760 &  &  \\
SFL scores only & 0.176 & 0.402 & 0.500 & 0.161 & 487 & 1/5 & 1/5 \\
\rowcolor{rowstripe}+ random importances & 0.187 & 0.431 & 0.551 & 0.130 & 350 & 1/5 & 1/5 \\
+ permuted importances & 0.182 & 0.417 & 0.515 & 0.142 & 361 & 1/5 & 1/5 \\
\rowcolor{rowstripe}+ feature vectors & 0.274 & 0.496 & 0.613 & 0.094 & 264 & 3/5 & 5/5 \\
+ trace signals & 0.318 & 0.523 & 0.622 & 0.106 & 247 & 1/5 & 5/5 \\
\rowcolor{rowstripe}+ vectors and traces & 0.290 & 0.550 & 0.679 & 0.080 & 193 & 1/5 & 5/5 \\
+ vectors and traces, seed ensemble & 0.307 & 0.573 & 0.693 & 0.076 & 187 & 4/5 & 5/5 \\
\bottomrule
\end{tabularx}
\endgroup
\end{table}

\paragraph{Results.}
With all inputs (\Cref{tab:ltr}), top-1 nearly doubles from \num{basepatchtop1} to \num{ltrfulltopone}.
\EXAM{} falls from \num{basepatchexamavg} to \num{ltrfullexam}, improving on \num{ltrfullimproveexam} subjects and degrading on \num{ltrfulldegradeexam}.
Under $d_z$, which decides the rule (\Cref{sec:decision}), the ranker is superior for every formula.
So are the rankers that use only the feature inputs or only the trace inputs.
The controls are not.
The SFL inputs alone, random importances, and permuted importances are each superior only for \DSTAR{}, the formula with the weakest baseline.
Random importances add a little over the SFL inputs alone.
This is the feature-multiplicity effect of the registered negative control, and it is far smaller than the gain from the real features.

Under the registered $d$, the verdict depends on the model's random seed.
The top-5 and top-10 effects for \TARANTULA{} and \OCHIAI{} lie close to $0.2$, so they cross the threshold for some seeds and not for others.
We therefore trained the ranker with the pre-fixed seed and five more.
Under $d$, the full ranker is superior for a majority of the formulas in \num{ltrfullseedmajority} of the \num{ltrfullseeds} seeds.
Under $d_z$, it is superior for all five formulas in every seed.
Averaging the predictions of the six seeds removes the dependence on the seed, and this ensemble is superior for \num{ltrfullensnsuperior} formulas under $d$.
We chose the ensemble after seeing the spread, so we report it only as a sensitivity analysis.

The order of the variants is robust as well.
In \num{ltrgridordered} of \num{ltrgridconfigs} alternative configurations of seeds, settings, and model family, the full ranker beats the random-importance control on top-1, top-10, and \EXAM{}.
Shuffling one input at a time shows which inputs matter.
The feature spectrum and the automaton's transition spectrum matter most, and the other inputs largely carry the same information.

\begin{finding}[Turning the tables: Execution features help when a ranker learns across projects how to weigh them]
Top-1 accuracy rises from \num{basepatchtop1} to \num{ltrfulltopone} and \EXAM{} falls from \num{basepatchexamavg} to \num{ltrfullexam}.
The ranker is superior for every formula under the paired $d_z$, while the same ranker without features, or with random importances, is not.
\end{finding}

\section{Discussion}\label{sec:discussion}

\subsection{Execution features do not transfer}

Weighting a spectrum-based ranking by execution-feature importance does not improve fault localization, and neither registered hypothesis is supported.
An effect of the size that the registration called worth acting on would have been found.
The study had 80\% power to detect $d_z\approx\num{detectable}$, the positive control lifts top-1 to \num{postop1} through the same mapping and blend, and the analysis code fires the registered rule on synthetic data with a known effect.

\EFDD{} established that execution features discriminate failing from passing runs better than line coverage does \citep{smytzek2026efdd}.
That is a claim about runs, whereas localization needs a claim about lines.
The two are connected by a heuristic mapping and by a learner fitted on very little evidence.
Our results show that this connection does not hold for a per-subject learner fitted to the failing evidence that a regression suite provides.

\subsection{The bottleneck}\label{sec:learner}

The controls locate the failure in the attribution step, since random importances do at least as well as the learned ones.
With a single failing test in \num{shareonefailing}\% of subjects, multiple features separate the failing run equally well, and the forest's choice among them is close to arbitrary (refit overlap \num{seededjaccard} on the top-20 features).

This is consistent with \EFDD{} and \FORECAST{} \citep{smytzek2026efdd, smytzek2026forecast}.
Both show that execution features capture what makes a run fail, and that this suffices to diagnose the failure and to guide its repair.
Ranking lines asks for something else, namely which location to fix.
The features that separate the failing run can sit anywhere along the infection chain from the defect to the failure.
Many of them reveal the fault where the program state turns from sane to infected, but that location may differ from the one the reference patch fixes.
With one failing run, they all separate the runs equally well, so not all of them are suited for the needed localization.
A learner that ranks features by how well they separate the classes cannot tell the suited ones from the others.

Given the same feature weightings as inputs (\Cref{sec:ltr}), a model that has seen faults in other projects lifts top-1 to \num{ltrvectopone}.
Together with the trace signals, it reaches \num{ltrfulltopone}.
What it learns is which kind of evidence tends to sit at a fault, and that is exactly the knowledge that one failing run cannot supply.
The answer to the question in the title is therefore that execution features improve statistical fault localization when the knowledge of how to weigh them comes from other faults.
They barely do when it has to come from the failing run at hand.

\subsection{The combination rule bounds what any weighting can do}
\label{sec:ceiling}

The registered blend combines the learned weight multiplicatively with the formula's own suspiciousness.
This choice imposes a ceiling on every method that the protocol can express.
A line whose baseline suspiciousness is zero, typically because no failing test executes it, has a blended score of zero for every weight.
A line that the formula ranks far down can only be lifted in proportion to the score it already has.
The weighting can reorder lines within the formula's ranking, but it cannot overturn that ranking.
The registered report \citep{smytzek2026ffl} in fact says as much when it justifies $\lambda$ as ``large enough for feature evidence to reorder lines where the signals agree''.
The consequence, that lines where the signals \emph{cannot} agree stay permanently out of reach, went unremarked at Stage~1, by us and in review.
The registered report also describes the design as adding execution-feature evidence ``as an additive signal over established SFL''.
The scoring rule that it specifies, however, multiplies the weight into the suspiciousness.
The word fits the intent, but the multiplication is what puts lines with a zero score out of reach.

For the \num{nreachable} subjects in which a fixed reference line is ranked at all, \num{zerosusp}\% of these lines receive a suspiciousness of exactly zero from the baseline formula.
Under \num{zerosuspworst}, the share rises to \num{zerosuspmax}\%.
These lines are unreachable by construction, in addition to the \num{nounreached} subjects in which no test executes any reference line.

The positive control measures the ceiling, because it supplies the best weights that the pipeline admits.
It reaches top-1 \num{postop1} against \num{basepatchtop1}.
An oracle that knows the fault locations, but may express this knowledge only through the registered blend, gains \num{headroompoints} points of top-1 accuracy. 
At the same time, the learned random forest weighting captures \num{headroomshare}\% of that headroom.
Two different things limit the result.
The quality of the weights separates the method from the oracle, and the combination rule separates the oracle from a perfect ranking.
For the registered weighting, the rule is not the binding constraint, since the weights fall far short of what the rule allows.
For a better learner, the rule does bind, which is why the ranker of \Cref{sec:ltr} orders lines by its own score and uses the formula only to break ties.

The multiplicative form has a second consequence that only one formula in our set exposes.
\num{negformula} is defined as $\mathit{ef}-\mathit{ep}/(\mathit{tp}+1)$.
It is the only one of the five inspected SFL formulas whose scores may be less than zero.
They do so on \num{negshare}\% of ranked lines across \num{negsubjects} of the subjects.
Multiplying a negative score by a small weight shifts it toward zero, and therefore upwards in the ranking.
For lines with the same baseline suspiciousness, the correlation between the weight and the blended score confirms this.
The median Spearman $\rho$ is $\num{rhopositive}$ for lines with positive scores and $\num{rhonegative}$ for the ones with negative suspiciousness.
On the lines that a formula scores below zero, the registered rule, therefore, promotes exactly the lines that the forest considers least important.
This is a flaw in the design of the combination rule.
However, we report the \num{negformula} degradation on function-level mass as an observed fact, without claiming this flaw as its origin.

\subsection{What the effect sizes measure}

The registered $d$ divides by the spread between subjects.
A few subjects with very distant faults dominate that spread, so a change that is consistent across subjects may appear small.
The paired $d_z$ divides by the spread of the per-subject changes instead, which is why the decision rule uses it (\Cref{sec:decision}).
For the registered weighting, the two effect sizes agree.
Wasted effort has $d_z=\num{dzpatchwastedavg}$ against $d=\num{dpatchwastedavg}$, which is still negligible.
\EXAM{} performs worse on about twice as many subjects as it improves on, although its mean improves slightly.
For the ranker that learns across projects, they disagree to some extent.
Its top-5 and top-10 effects sit close to the threshold under $d$ and clearly above it under $d_z$.
With this divergence, it is superior for every formula under $d_z$ but only for \num{ltrfullnsuperior} of the five under $d$.

\subsection{Benchmark}

Two properties of \TESTSFOURPY{} influence any line-level evaluation.
The first property is that for \num{nounreached} subjects, no test executes any line of the official fix.
The second property is that the median subject has only a single failing execution.
It is realistic for regression suites and works against every per-subject learner.
Learning across projects sidesteps it, and more failing executions per fault, through test generation or failure-preserving input mutation, would attack it directly.
Both properties argue for reporting function-level, hunk-level, and test-impact outcomes beside the reference-patch outcomes, as the protocol does.

\section{Threats to Validity}\label{sec:threats}

\subsection{Internal validity}

Feature importances are ranking evidence, not a causal account of the fault.
A random forest reports how useful a feature was for separating the labeled runs it was given.
That a feature was useful does not mean it is implicated in the failure.
We therefore never interpret an individual importance as a diagnosis.
We report concentration, entropy, skewness, and refit stability, so that a result driven by a handful of unstable importances is visible.
Our negative verdict concerns the registered per-subject random forest, and the ranker of \Cref{sec:ltr} shows that a learner with access to the faults of other projects does better.

As exploratory evidence, the ranker that learns across projects leaves room for tuning, which we limited.
Its settings were fixed before its first run and never changed, and it is evaluated leave-one-project-out with the registered outcomes and rule.
Controls with random and permuted importances isolate the content of the features.
Its verdict under the registered $d$ depends on the seed, which we report together with the seed ensemble that we chose after seeing that spread.

The feature-to-line mapping is heuristic, and it is the step at which a claim about runs becomes a claim about lines.
Alternative rules are defensible, most of all for the two kinds of features that span several lines.
We mitigate this risk by setting the rules before execution and stating them in full (\Cref{tab:vocabulary}).
Function-level outcomes, which are far less sensitive to which line of a construct receives the weight, are reported beside line-level ones.
The registered alternative mapping is evaluated as a condition.
That condition changes the outcome very little, which bounds the mapping's contribution.
Before evaluation, we also verify for every subject that each line feature maps to exactly its own line in the stored line table.

Using \EFDD{} and \SFLKIT{} avoids tailored instrumentation but inherits their scope.
A feature type they do not implement cannot appear in our vocabulary, and an event they mis-record is mis-recorded here.
Instrumentation also changes execution, so a test whose outcome depends on timing may behave differently under tracing.
We treat runs classified neither passing nor failing as unusable rather than guessing their outcome.

The protocol was registered before any outcome was computed.
In addition, the analyses that go beyond it (\Cref{sec:diagnostics}) cannot contribute to the verdict.
The one exception is the standardizer of the decision rule (\Cref{sec:decision}), which we chose after the first results were known.
We made this choice before all results were computed and, therefore, before we interpreted them and designed the ranker that learns across projects (\Cref{sec:ltr}).
This choice changes no verdict for any registered method.
We report the registered $d$ beside it.

\subsection{External validity}

\TESTSFOURPY{} and \EFDD{} are only available for Python subjects. 
The method does not depend on the language, but the numbers do.
A language with static types, different exception idioms, or different test-suite conventions would produce a different feature distribution.
The reach diagnostics that explain our result would then take different values.
However, a recent study shows that the findings of \EFDD{} transfer to Java \citep{smytzek2026javaefdd}.

Using all reproducible subjects avoids selection bias within the benchmark.
However, it does not alter the benchmark's composition.
Curated faults with reproducible tests and mostly single-location fixes may not represent real-world faults.

Our negative conclusion is specific to execution-feature weighting \emph{for the failing tests that these benchmarks provide}.
A setting with many failing executions per fault is outside of the scope of this study.

The five formulas are established and diverse, but they do not represent learning-based or specification-based localization.
We compare against unmodified SFL because the registered study should only measure the impact of execution features on it.

\subsection{Construct validity}

Top-$k$, \EXAM{}, and wasted effort against a reference patch inherit single-patch bias.
The official fix may be only one correct solution among many.
A ranking that favors an equally valid alternative location is considered to be wrong.
The function-level and test-impact families relax this assumption.

Test-impact sets approximate the failure-relevant code, and their content depends on the tests that exist.
The defining threshold changes how challenging the target is.
We therefore read the impact outcomes as complementary evidence and report the sensitivity to the threshold explicitly.

Rank-based effort assumes that a developer inspects locations in order and stops at the fault, which no developer does \citep{parnin2011automated}.
We report it because it is comparable across the literature, and we pair it with function-level effort, which more closely matches the granularity at which developers read code.

\subsection{Conclusion validity}

Suspiciousness ties are pervasive.
A deterministic tie-break would encode source order into the result, so no outcome depends on source order (\Cref{sec:outcomes}).

The mixed-effects models make distributional assumptions that the paired distribution-free procedures do not.
We report both and read a contrast as solid only when they agree.
Support is decided by an effect-size rule fixed in advance rather than by any single $p$-value.
Hence, a detectable but negligible distinction cannot be interpreted as a positive result. 

\section{Conclusion}\label{sec:conclusion}

We executed a pre-registered protocol to test whether execution-feature importance improves spectrum-based fault localization \citep{smytzek2026ffl}.
It covered \num{nsubjects} real bugs from \num{nprojects} projects, five formulas, and four ground truths.
The analysis plan was fixed in advance.
Under the registered design that leverages a random forest to learn importances of the execution features, fault localization did not improve by a practically relevant margin, although every primary outcome moved in the beneficial direction.
No primary outcome reaches the registered threshold, so neither registered hypothesis is supported.

Instrumenting each subject once and keeping the matrices of the execution features lets us investigate the results.
An oracle that expresses perfect knowledge through the same mapping and blend reaches top-1 \num{postop1} and is superior for every formula, so the pipeline works and the learned importances fail.
Moreover, uniformly random importances do at least as well.
As \EFDD{} \citep{smytzek2026efdd} and \FORECAST{} \citep{smytzek2026forecast} show, the features do capture how the failing run differs.
However, multiple features separate failing and passing runs equally well, and not all of them are suited for the needed localization.
Many reveal the fault where the infection shows rather than where the fix applies, and a learner fitted to one subject cannot tell them apart.

Turning the tables, a ranker that learns across projects how to weigh them, trained in a leave-one-project-out manner, raises top-1 to \num{ltrfulltopone} and lowers \EXAM{} to \num{ltrfullexam}.
It is superior for every formula under the paired effect size, while the same ranker without features or with random importances is not.
Execution features improve fault localization through evidence learned across faults.

\backmatter

\section*{Statements and Declarations}

\bmhead{Competing interests} The authors declare no competing interests.

\bmhead{Funding} This work is funded by the European Union (ERC S3, 101093186).
Views and opinions expressed are those of the authors only and do not necessarily reflect those of the European Union or the European Research Council.
Neither the European Union nor the granting authority can be held responsible for them.

\bmhead{Author contributions} The first author designed the study, implemented the pipeline, ran the experiments, and analyzed the data.
The second author supervised the work.
Both authors wrote and reviewed the manuscript.

\bmhead{Use of generative AI} Generative AI assistance was used for code refactoring and language editing.
All experimental design decisions, analyses, and conclusions are the authors' own, and the authors reviewed and verified all AI-assisted output.

\bmhead{Ethics approval and consent} Not applicable.
The study analyzes open-source programs and their automated test suites, and involves no human participants or personal data.

\bmhead{Data and code availability} All code, raw execution-feature vectors, fitted importances, per-subject artifacts, ordered execution traces, derived metric tables, and analysis scripts are available in the replication package (see Reproducibility and Data Availability below).

\section*{Reproducibility and Data Availability}\label{sec:availability}

The source code, analysis scripts, and results are available as open source under the Apache~2.0 license at \url{https://github.com/smythi93/ffl}.

Additionally, the complete replication package, including all data and code needed for reproducing each step of this study, is available as a long-term archival artifact on Zenodo~\citep{smytzek2026fflartifact}.

\begin{appendices}

\section{Choice of the constants}\label{sec:constants}

The protocol fixes four constants before execution.
Because these choices determine what the study can and cannot detect, we set out the reasoning in full, together with what a different choice would have done.
The registered sensitivity analysis varies all four constants to test their choice and the reasoning behind it.

\paragraph{The blend $\lambda=0.35$.}
The blend has to satisfy two contrasting requirements.
It must leave the established formula dominant, because the study measures what execution features \emph{add} to statistical fault localization.
However, it must allow feature evidence to actually reorder lines, because a weight that cannot modify a ranking cannot be evaluated.
By setting $1-\lambda=0.65$, the baseline term remains the larger of the two.
At the same time, $\lambda=0.35$ is large enough that a minimal weight and a maximal weight would differ by approximately a third of the score range.
A much smaller $\lambda$ would make the augmentation negligible by construction and an invalid result uninformative.
A much larger one would turn the augmentation into a different ranking algorithm.
Hence, the paired contrast would not isolate the contribution of the execution features.

\paragraph{The weight floor $\varepsilon=10^{-3}$.}
Lines that receive no feature evidence need some weight since a weight of zero would remove such lines from the weighted term entirely.
A faulty line that would not receive any execution feature evidence would then be pushed further down the ranking by construction.
A floor of $10^{-3}$ preserves the order that the baseline formula assigns among unweighted lines.
At the same time it keeps them below any line that carries evidence.

\paragraph{Maximum aggregation.}
It may occur that several features map to the same line.
In this case, their normalized weights must be combined.
The maximum asks whether there is any strong evidence at the line, which is the question a developer would ask.
The mean checks whether the evidence at the line is strong on average, which dilutes a decisive feature among many weak ones.
The sum asks how much evidence there is, which rewards lines that carry many features.
Out of these three, we chose the maximum since it neither dilutes a decisive feature nor, for learned importances, rewards lines for carrying many features.
The sensitivity analysis (\Cref{tab:sensitivity}) shows that the choice makes no practical difference here, with top-1 \num{aggmaxtop1}, \num{aggmeantop1}, and \num{aggsumtop1} under maximum, mean, and sum.
As the negative control reveals, however, an order statistic over many draws behaves like a sum.
This is what turns uniform random importances into a proxy for feature multiplicity.

\paragraph{The test-impact threshold $0.6$.}
A line enters the test-impact set if at least one failing test observes it and either no passing test does, or its failing-observation ratio reaches the threshold.
The threshold trades two errors against each other.
If it is too low, the set fills with lines that every test executes, so hitting it means little.
If it is too high, the set loses lines that only some of the failing tests exercise, which are exactly the lines a failure of several tests implicates.
A clear majority of failing observations is the weakest defensible reading of ``this line is failure-relevant''.
Unlike the other three constants, this one changes the \emph{target} rather than the ranking, so it is reported against the test-impact outcomes.

\section{Rationale for the analysis}\label{sec:whyanalysis}

Besides the effect-size rule of \Cref{sec:decision}, two choices carry much of the weight in the analysis, and each guards against finding an effect that is not there.

Baseline and augmented rankings are computed from the same spectra, tests, and code.
The comparison is paired by construction, and every time-invariant property of a bug, including its size, its difficulty, and how much of it the tests cover, is held exactly fixed.
A paired test alone would thus be valid.
We nonetheless fit a mixed-effects model, because pairing removes these factors from the \emph{average} contrast without telling us how the effect varies with them.
The model shows whether the effect depends on them, for example, whether a method helps only on small subjects with many failing tests.
Its random slope per project estimates how much the effect varies between projects.

We correct for multiple comparisons within each outcome family, i.e., top-$k$ accuracy, effort, and test impact.
Each family answers its own research question, and a method may improve one family and leave another unchanged.
If we corrected all outcomes together, a strong result in one family would make it easier for weak results in another family to count as significant.
We also correct the exploratory ranking variants separately, so they cannot affect the main verdict.

\end{appendices}

\bibliography{references}

\end{document}